\documentclass[preprint,5p,times,twocolumn,nonatbib]{elsarticle}

\journal{Safety Science}

\usepackage{amssymb,amsmath,amsfonts}
\usepackage{algorithmic}
\usepackage{graphicx}
\usepackage{hyperref} 

\usepackage{textcomp}

\usepackage{verbatim}
\usepackage[style=ieee,
bibstyle=ieee,
backend=biber,
minnames=1,
maxcitenames=2,
maxbibnames=6,
doi=true,
isbn=false,
url=false,
date=year,
]{biblatex} 

\AtEveryBibitem{\clearlist{language}}
\usepackage{xpatch}
\usepackage{xstring}

\DeclareSourcemap{
	\maps[datatype=bibtex]{
		\map[overwrite, foreach={booktitle,journaltitle,eventtitle,series,publisher,institution,type}]{
			\step[fieldsource=\regexp{$MAPLOOP}, match={XXX}, replace={Acad. Serbe Sci. Arts Glas Cl. Sci. Tech.}]
			\step[fieldsource=\regexp{$MAPLOOP}, match={XXX}, replace={Acta Acust.}]
			\step[fieldsource=\regexp{$MAPLOOP}, match={XXX}, replace={Acta Astron. (Poland)}]
			\step[fieldsource=\regexp{$MAPLOOP}, match={XXX}, replace={Acta Astron. Sin. (China)}]
			\step[fieldsource=\regexp{$MAPLOOP}, match={XXX}, replace={Acta Astronaut. (U.K.)}]
			\step[fieldsource=\regexp{$MAPLOOP}, match={XXX}, replace={Acta Astrophys. Sin. (China)}]
			\step[fieldsource=\regexp{$MAPLOOP}, match={XXX}, replace={Acta Autom. Sin.}]
			\step[fieldsource=\regexp{$MAPLOOP}, match={XXX}, replace={Acta Cienc. Indica Math.}]
			\step[fieldsource=\regexp{$MAPLOOP}, match={XXX}, replace={Acta Cienc. Indica Phys.}]
			\step[fieldsource=\regexp{$MAPLOOP}, match={XXX}, replace={Acta Crystallogr. A, Found. Crystallogr.}]
			\step[fieldsource=\regexp{$MAPLOOP}, match={XXX}, replace={Acta Crystallogr. B, Struct. Sci.}]
			\step[fieldsource=\regexp{$MAPLOOP}, match={XXX}, replace={Acta Crystallogr. C, Cryst. Struct. Commun.}]
			\step[fieldsource=\regexp{$MAPLOOP}, match={XXX}, replace={Acta Electron. (France)}]
			\step[fieldsource=\regexp{$MAPLOOP}, match={XXX}, replace={Acta Cyberno}]
			\step[fieldsource=\regexp{$MAPLOOP}, match={XXX}, replace={Acta Electron. Sin. (China)}]
			\step[fieldsource=\regexp{$MAPLOOP}, match={XXX}, replace={Acta Geod. Geophys. Montan. Hung.}]
			\step[fieldsource=\regexp{$MAPLOOP}, match={XXX}, replace={Acta Geophys. Pol.}]
			\step[fieldsource=\regexp{$MAPLOOP}, match={XXX}, replace={Acta Geophys. Sin. (China)}]
			\step[fieldsource=\regexp{$MAPLOOP}, match={XXX}, replace={Acta Geophys. Sin. (USA)}]
			\step[fieldsource=\regexp{$MAPLOOP}, match={XXX}, replace={Acta Metall.}]
			\step[fieldsource=\regexp{$MAPLOOP}, match={XXX}, replace={Acta Mex. Cienc. Tecnol.}]
			\step[fieldsource=\regexp{$MAPLOOP}, match={XXX}, replace={Acta Phys. Hung.}]
			\step[fieldsource=\regexp{$MAPLOOP}, match={XXX}, replace={Acta Phys. Pol. A}]
			\step[fieldsource=\regexp{$MAPLOOP}, match={XXX}, replace={Acta Phys. Pol. B}]
			\step[fieldsource=\regexp{$MAPLOOP}, match={XXX}, replace={Acta Phys. Sin.}]
			\step[fieldsource=\regexp{$MAPLOOP}, match={XXX}, replace={Acta Phys. Slovaca}]
			\step[fieldsource=\regexp{$MAPLOOP}, match={XXX}, replace={Acta Politec. Mex.}]
			\step[fieldsource=\regexp{$MAPLOOP}, match={XXX}, replace={Acta Polytech. Scand. Appl. Phys. Ser.}]
			\step[fieldsource=\regexp{$MAPLOOP}, match={XXX}, replace={Acta Polytech. Scand. Chem. Technol.}]
			\step[fieldsource=\regexp{$MAPLOOP}, match={XXX}, replace={Metall. Ser.}]
			\step[fieldsource=\regexp{$MAPLOOP}, match={XXX}, replace={Acta Polytech. Scand. Electr. Eng. Ser.}]
			\step[fieldsource=\regexp{$MAPLOOP}, match={XXX}, replace={Acta Polytech. Scand. Math. Comput. Sci. Ser.}]
			\step[fieldsource=\regexp{$MAPLOOP}, match={XXX}, replace={Acta Polytech. Scand. Mech. Eng. Ser.}]
			\step[fieldsource=\regexp{$MAPLOOP}, match={XXX}, replace={Acta Seismol. Sin.}]
			\step[fieldsource=\regexp{$MAPLOOP}, match={XXX}, replace={Acta Tech. Acad. Sci. Hung.}]
			\step[fieldsource=\regexp{$MAPLOOP}, match={XXX}, replace={Acta Tech. CSAV}]
			\step[fieldsource=\regexp{$MAPLOOP}, match={XXX}, replace={Acustica}]
			\step[fieldsource=\regexp{$MAPLOOP}, match={XXX}, replace={(AEU) Arch. Elektr. Ubertragung}]
			\step[fieldsource=\regexp{$MAPLOOP}, match={XXX}, replace={Akust. Zh.}]
			\step[fieldsource=\regexp{$MAPLOOP}, match={XXX}, replace={Algorithmica}]
			\step[fieldsource=\regexp{$MAPLOOP}, match={XXX}, replace={Alta Freq.}]
			\step[fieldsource=\regexp{$MAPLOOP}, match={XXX}, replace={An. Acad. Bras. Cienc.}]
			\step[fieldsource=\regexp{$MAPLOOP}, match={XXX}, replace={An. Fis.}]
			\step[fieldsource=\regexp{$MAPLOOP}, match={XXX}, replace={An. Mec. Electr.}]
			\step[fieldsource=\regexp{$MAPLOOP}, match={XXX}, replace={Angew. Inform.}]
			\step[fieldsource=\regexp{$MAPLOOP}, match={XXX}, replace={Ann. Inst. Henri Poincare Phys. Theor.}]
			\step[fieldsource=\regexp{$MAPLOOP}, match={XXX}, replace={Ann. Soc. Sci. Brux. I, Sci. Math. Astron. Phys.}]
			\step[fieldsource=\regexp{$MAPLOOP}, match={XXX}, replace={Arch. Elektr. Uebertrag. (AEU)}] 
			\step[fieldsource=\regexp{$MAPLOOP}, match={XXX}, replace={Arch. Elektron. Uebertrag. Tech.}] 
			\step[fieldsource=\regexp{$MAPLOOP}, match={XXX}, replace={Arch. Elektrotech. (Poland)}]
			\step[fieldsource=\regexp{$MAPLOOP}, match={XXX}, replace={Arch. Elektrotech. (Germany)}]
			\step[fieldsource=\regexp{$MAPLOOP}, match={XXX}, replace={Ark. Fys. Semin. Trondheim}]
			\step[fieldsource=\regexp{$MAPLOOP}, match={XXX}, replace={Astrofizika}]
			\step[fieldsource=\regexp{$MAPLOOP}, match={XXX}, replace={Astron. Nachr. (Germany)}]
			\step[fieldsource=\regexp{$MAPLOOP}, match={XXX}, replace={Astron. Tidsskr.}]
			\step[fieldsource=\regexp{$MAPLOOP}, match={XXX}, replace={Astron. Vestn.}]
			\step[fieldsource=\regexp{$MAPLOOP}, match={XXX}, replace={Astron. Zh.}]
			\step[fieldsource=\regexp{$MAPLOOP}, match={XXX}, replace={Atomwirtsch.-Atomtech.}]
			\step[fieldsource=\regexp{$MAPLOOP}, match={XXX}, replace={Atti Accad. Sci. Ist. Bologna CI. Sci. Fis.}]
			\step[fieldsource=\regexp{$MAPLOOP}, match={XXX}, replace={Rend. XIII}]
			\step[fieldsource=\regexp{$MAPLOOP}, match={XXX}, replace={Atti Accad. Sci. Torino I, CI. Sci. Fis. Math. Nat.}]
			\step[fieldsource=\regexp{$MAPLOOP}, match={XXX}, replace={Autom. Strum.}]
			\step[fieldsource=\regexp{$MAPLOOP}, match={XXX}, replace={Autom. Tech. Prax.}]
			\step[fieldsource=\regexp{$MAPLOOP}, match={XXX}, replace={Automatica}]
			\step[fieldsource=\regexp{$MAPLOOP}, match={XXX}, replace={Automatie}]
			\step[fieldsource=\regexp{$MAPLOOP}, match={XXX}, replace={Automatika}]
			\step[fieldsource=\regexp{$MAPLOOP}, match={XXX}, replace={Automatisierungstechnik}]
			\step[fieldsource=\regexp{$MAPLOOP}, match={XXX}, replace={Automatizace}]
			\step[fieldsource=\regexp{$MAPLOOP}, match={XXX}, replace={Automedica}]
			\step[fieldsource=\regexp{$MAPLOOP}, match={XXX}, replace={Avtom. Telemekh.}]
			\step[fieldsource=\regexp{$MAPLOOP}, match={XXX}, replace={Avtom. Vychisl. Tekh.}]
			\step[fieldsource=\regexp{$MAPLOOP}, match={XXX}, replace={Avtomatika}]
			\step[fieldsource=\regexp{$MAPLOOP}, match={XXX}, replace={Avtometriya}]
			\step[fieldsource=\regexp{$MAPLOOP}, match={ATZelektronik worldwide}, replace={ATZelektron. worldw.}]
			\step[fieldsource=\regexp{$MAPLOOP}, match={XXX}, replace={Ber. Bunsenges. Phys. Chem.}]
			\step[fieldsource=\regexp{$MAPLOOP}, match={XXX}, replace={Biofizika}]
			\step[fieldsource=\regexp{$MAPLOOP}, match={XXX}, replace={Biometrika}]
			\step[fieldsource=\regexp{$MAPLOOP}, match={XXX}, replace={Boll. Geofis. Teor. Appl.}]
			\step[fieldsource=\regexp{$MAPLOOP}, match={XXX}, replace={Bull. Acad. Serbe Sci. Arts cl. Sci. Tech.}]
			\step[fieldsource=\regexp{$MAPLOOP}, match={XXX}, replace={Bull. Annu. Soc. Suisse Chronom. Lab. Suisse.}]
			\step[fieldsource=\regexp{$MAPLOOP}, match={XXX}, replace={Rech. Horlog.}]
			\step[fieldsource=\regexp{$MAPLOOP}, match={XXX}, replace={Bull. Cl. Sci. Acad. R. Belg.}]
			\step[fieldsource=\regexp{$MAPLOOP}, match={XXX}, replace={Bull. Dir. Etud. Rech. A}]
			\step[fieldsource=\regexp{$MAPLOOP}, match={XXX}, replace={Bull. Dir. Etud. Rech. B}]
			\step[fieldsource=\regexp{$MAPLOOP}, match={XXX}, replace={Bull. Dir. Etud. Rech. C}]
			\step[fieldsource=\regexp{$MAPLOOP}, match={XXX}, replace={Bull. Liaison Rech. Inform. Autom.}]
			\step[fieldsource=\regexp{$MAPLOOP}, match={XXX}, replace={Bur. Etud. Autom.}]
			\step[fieldsource=\regexp{$MAPLOOP}, match={XXX}, replace={CFI-Ceram. Forum Int.-Ber. Dtsch. Keram. Ges.}]
			\step[fieldsource=\regexp{$MAPLOOP}, match={XXX}, replace={Chem. Scr.}]
			\step[fieldsource=\regexp{$MAPLOOP}, match={XXX}, replace={Ciel Terre}]
			\step[fieldsource=\regexp{$MAPLOOP}, match={XXX}, replace={Cybernetica}]
			\step[fieldsource=\regexp{$MAPLOOP}, match={XXX}, replace={Deut. Hydrogr. Z.}]
			\step[fieldsource=\regexp{$MAPLOOP}, match={XXX}, replace={Dokl. Akad. Nauk SSSR}]
			\step[fieldsource=\regexp{$MAPLOOP}, match={XXX}, replace={Electroacoustique}]
			\step[fieldsource=\regexp{$MAPLOOP}, match={XXX}, replace={Electrochim. Acta}]
			\step[fieldsource=\regexp{$MAPLOOP}, match={XXX}, replace={Electrochim. Metal.}]
			\step[fieldsource=\regexp{$MAPLOOP}, match={XXX}, replace={Elektor Electron.}]
			\step[fieldsource=\regexp{$MAPLOOP}, match={XXX}, replace={Elektr. Bahnen}]
			\step[fieldsource=\regexp{$MAPLOOP}, match={XXX}, replace={Elektr. Energ.-Tech.}]
			\step[fieldsource=\regexp{$MAPLOOP}, match={XXX}, replace={Elektr. Masch.}]
			\step[fieldsource=\regexp{$MAPLOOP}, match={XXX}, replace={Elektr. Stn.}]
			\step[fieldsource=\regexp{$MAPLOOP}, match={XXX}, replace={Elektrichestvo}]
			\step[fieldsource=\regexp{$MAPLOOP}, match={XXX}, replace={Elektrie}]
			\step[fieldsource=\regexp{$MAPLOOP}, match={XXX}, replace={Elektrizitaetswirtschaft}]
			\step[fieldsource=\regexp{$MAPLOOP}, match={XXX}, replace={Elektro}]
			\step[fieldsource=\regexp{$MAPLOOP}, match={XXX}, replace={Elektro-Anz.}]
			\step[fieldsource=\regexp{$MAPLOOP}, match={XXX}, replace={Elektro-Jahr}]
			\step[fieldsource=\regexp{$MAPLOOP}, match={XXX}, replace={Elektrokhimiya}]
			\step[fieldsource=\regexp{$MAPLOOP}, match={XXX}, replace={Elektron}]
			\step[fieldsource=\regexp{$MAPLOOP}, match={XXX}, replace={Elektron Int.}]
			\step[fieldsource=\regexp{$MAPLOOP}, match={XXX}, replace={Elektron. Entwitkl.}]
			\step[fieldsource=\regexp{$MAPLOOP}, match={XXX}, replace={Elektron. Ind}]
			\step[fieldsource=\regexp{$MAPLOOP}, match={XXX}, replace={Elektron. J.}]
			\step[fieldsource=\regexp{$MAPLOOP}, match={XXX}, replace={Elektron. Prax.}]
			\step[fieldsource=\regexp{$MAPLOOP}, match={XXX}, replace={Elektron. Tekh.}]
			\step[fieldsource=\regexp{$MAPLOOP}, match={XXX}, replace={Elektronica}]
			\step[fieldsource=\regexp{$MAPLOOP}, match={XXX}, replace={Elektronik}]
			\step[fieldsource=\regexp{$MAPLOOP}, match={XXX}, replace={Elektronika}]
			\step[fieldsource=\regexp{$MAPLOOP}, match={XXX}, replace={Elektroniker}]
			\step[fieldsource=\regexp{$MAPLOOP}, match={XXX}, replace={Elektronikschau}]
			\step[fieldsource=\regexp{$MAPLOOP}, match={XXX}, replace={Elektrosvyaz}]
			\step[fieldsource=\regexp{$MAPLOOP}, match={XXX}, replace={Elektrotech. Cas.}]
			\step[fieldsource=\regexp{$MAPLOOP}, match={Elektrotechnik und Informationstechnik}, replace={Elektrotech. Inf. Tech.}]
			\step[fieldsource=\regexp{$MAPLOOP}, match={XXX}, replace={Elektrotech. Obz.}]
			\step[fieldsource=\regexp{$MAPLOOP}, match={XXX}, replace={Elektrotechniek}]
			\step[fieldsource=\regexp{$MAPLOOP}, match={XXX}, replace={Elektrotechnik (Czechoslovakia)}]
			\step[fieldsource=\regexp{$MAPLOOP}, match={XXX}, replace={Elektrotechnik (Switzerland)}]
			\step[fieldsource=\regexp{$MAPLOOP}, match={XXX}, replace={Elektrotechnik (Germany)}]
			\step[fieldsource=\regexp{$MAPLOOP}, match={XXX}, replace={Elektrotechnika}]
			\step[fieldsource=\regexp{$MAPLOOP}, match={XXX}, replace={Elektrotehnika, Zagreb}]
			\step[fieldsource=\regexp{$MAPLOOP}, match={XXX}, replace={Elektrotekhnika}]
			\step[fieldsource=\regexp{$MAPLOOP}, match={XXX}, replace={Elektroteknikeren}]
			\step[fieldsource=\regexp{$MAPLOOP}, match={XXX}, replace={Elektrowaerme Int. B.}]
			\step[fieldsource=\regexp{$MAPLOOP}, match={XXX}, replace={Elettrificazione}]
			\step[fieldsource=\regexp{$MAPLOOP}, match={XXX}, replace={Elettron. Oggi}]
			\step[fieldsource=\regexp{$MAPLOOP}, match={XXX}, replace={Elettron. Telecomun.}]
			\step[fieldsource=\regexp{$MAPLOOP}, match={XXX}, replace={Elettrotecnica}]
			\step[fieldsource=\regexp{$MAPLOOP}, match={XXX}, replace={Elek. Med Aktuell Elektron.}]
			\step[fieldsource=\regexp{$MAPLOOP}, match={XXX}, replace={Elteknik}]
			\step[fieldsource=\regexp{$MAPLOOP}, match={XXX}, replace={Energ. Atomtech.}]
			\step[fieldsource=\regexp{$MAPLOOP}, match={XXX}, replace={Energ. Elettr.}]
			\step[fieldsource=\regexp{$MAPLOOP}, match={XXX}, replace={Energetica}]
			\step[fieldsource=\regexp{$MAPLOOP}, match={XXX}, replace={Energetik}]
			\step[fieldsource=\regexp{$MAPLOOP}, match={XXX}, replace={Energetika}]
			\step[fieldsource=\regexp{$MAPLOOP}, match={XXX}, replace={Energetyka}]
			\step[fieldsource=\regexp{$MAPLOOP}, match={XXX}, replace={Energia Nuclear}]
			\step[fieldsource=\regexp{$MAPLOOP}, match={XXX}, replace={Energie Technik (Germany)}]
			\step[fieldsource=\regexp{$MAPLOOP}, match={XXX}, replace={Energie Technik (Switzerland)}]
			\step[fieldsource=\regexp{$MAPLOOP}, match={XXX}, replace={Entropie}]
			\step[fieldsource=\regexp{$MAPLOOP}, match={XXX}, replace={ETZ}]
			\step[fieldsource=\regexp{$MAPLOOP}, match={XXX}, replace={ETZ Arch.}]
			\step[fieldsource=\regexp{$MAPLOOP}, match={XXX}, replace={Feingeraetetechnik}]
			\step[fieldsource=\regexp{$MAPLOOP}, match={XXX}, replace={Feinw. Tech. Messtech.}]
			\step[fieldsource=\regexp{$MAPLOOP}, match={XXX}, replace={Fert. Tech. Betr.}]
			\step[fieldsource=\regexp{$MAPLOOP}, match={XXX}, replace={Fis. Tecnol.}]
			\step[fieldsource=\regexp{$MAPLOOP}, match={XXX}, replace={Fiz. Khim. Obrab. Mater.}]
			\step[fieldsource=\regexp{$MAPLOOP}, match={XXX}, replace={Fiz. Met. Metalloved}]
			\step[fieldsource=\regexp{$MAPLOOP}, match={XXX}, replace={Fiz. Nizk. Temp.}]
			\step[fieldsource=\regexp{$MAPLOOP}, match={XXX}, replace={Fiz. Plazmy}]
			\step[fieldsource=\regexp{$MAPLOOP}, match={XXX}, replace={Fiz. Tekh. Poluprovodn.}]
			\step[fieldsource=\regexp{$MAPLOOP}, match={XXX}, replace={Fiz. Tverd. Tela}]
			\step[fieldsource=\regexp{$MAPLOOP}, match={XXX}, replace={Fiz.-Khim. Mekh. Mater.}]
			\step[fieldsource=\regexp{$MAPLOOP}, match={XXX}, replace={Fizika}]
			\step[fieldsource=\regexp{$MAPLOOP}, match={XXX}, replace={Forsch.-Ber. Landes Nordrh.-Westfal.}]
			\step[fieldsource=\regexp{$MAPLOOP}, match={XXX}, replace={Frequenz}]
			\step[fieldsource=\regexp{$MAPLOOP}, match={XXX}, replace={Fys. Tidsskr.}]
			\step[fieldsource=\regexp{$MAPLOOP}, match={XXX}, replace={G. Fis.}]
			\step[fieldsource=\regexp{$MAPLOOP}, match={XXX}, replace={Geliotekhnika}]
			\step[fieldsource=\regexp{$MAPLOOP}, match={XXX}, replace={Geochim. Cosmochim. Acta}]
			\step[fieldsource=\regexp{$MAPLOOP}, match={XXX}, replace={Haerterei-Tech. Mitt.}]
			\step[fieldsource=\regexp{$MAPLOOP}, match={XXX}, replace={Helv. Chim. Acta}]
			\step[fieldsource=\regexp{$MAPLOOP}, match={XXX}, replace={Helv. Med. Acta}]
			\step[fieldsource=\regexp{$MAPLOOP}, match={XXX}, replace={Helv. Phys. Acta}]
			\step[fieldsource=\regexp{$MAPLOOP}, match={XXX}, replace={Hochfreq. Electroakust}]
			\step[fieldsource=\regexp{$MAPLOOP}, match={XXX}, replace={Hoppe-Seylers Z. Physiol. Chem.}]
			\step[fieldsource=\regexp{$MAPLOOP}, match={XXX}, replace={Inf. Elektron.}]
			\step[fieldsource=\regexp{$MAPLOOP}, match={XXX}, replace={Inf. Elettron.}]
			\step[fieldsource=\regexp{$MAPLOOP}, match={XXX}, replace={Inform. Forsch. Entwickl.}]
			\step[fieldsource=\regexp{$MAPLOOP}, match={XXX}, replace={Inform. Spektrum}]
			\step[fieldsource=\regexp{$MAPLOOP}, match={XXX}, replace={Inform.-Fachber.}]
			\step[fieldsource=\regexp{$MAPLOOP}, match={XXX}, replace={Informatie}]
			\step[fieldsource=\regexp{$MAPLOOP}, match={XXX}, replace={Informatik}]
			\step[fieldsource=\regexp{$MAPLOOP}, match={XXX}, replace={Informatologia Yugosl.}]
			\step[fieldsource=\regexp{$MAPLOOP}, match={XXX}, replace={Informatyka}]
			\step[fieldsource=\regexp{$MAPLOOP}, match={XXX}, replace={Infowelt}]
			\step[fieldsource=\regexp{$MAPLOOP}, match={XXX}, replace={Ing. Electr. Mec.}]
			\step[fieldsource=\regexp{$MAPLOOP}, match={XXX}, replace={Ing. Mec. Electr.}]
			\step[fieldsource=\regexp{$MAPLOOP}, match={XXX}, replace={Ing.-Arch.}]
			\step[fieldsource=\regexp{$MAPLOOP}, match={XXX}, replace={Inzh.-Fiz. Zh.}]
			\step[fieldsource=\regexp{$MAPLOOP}, match={XXX}, replace={Izmer. Tekh.}]
			\step[fieldsource=\regexp{$MAPLOOP}, match={XXX}, replace={Izv. Akad. Nauk Arm. SSR Ser. Tekh. Nauk}]
			\step[fieldsource=\regexp{$MAPLOOP}, match={XXX}, replace={Izv. Akad. Nauk SSSR Energ. Transp.}]
			\step[fieldsource=\regexp{$MAPLOOP}, match={XXX}, replace={Izv. Akad. Nauk SSSR Fiz. Atmos. Okeana}]
			\step[fieldsource=\regexp{$MAPLOOP}, match={XXX}, replace={Izv. Akad. Nauk SSSR Fiz. Zemli}]
			\step[fieldsource=\regexp{$MAPLOOP}, match={XXX}, replace={Izv. Akad. Nauk SSSR Ser. Fiz.}]
			\step[fieldsource=\regexp{$MAPLOOP}, match={XXX}, replace={Izv. Vyssh. Uchebn. Zaved. Elektromekh.}]
			\step[fieldsource=\regexp{$MAPLOOP}, match={XXX}, replace={Izv. Vyssh. Uchebn. Zaved. Radioelektron.}]
			\step[fieldsource=\regexp{$MAPLOOP}, match={XXX}, replace={Izv. Vyssh. Uchebn. Zaved. Radiofiz.}]
			\step[fieldsource=\regexp{$MAPLOOP}, match={XXX}, replace={J. Chim Phys. Phys.-Chim Biol.}]
			\step[fieldsource=\regexp{$MAPLOOP}, match={XXX}, replace={Kernenergie}]
			\step[fieldsource=\regexp{$MAPLOOP}, match={XXX}, replace={Kerntechnik}]
			\step[fieldsource=\regexp{$MAPLOOP}, match={XXX}, replace={Khim. Fiz.}]
			\step[fieldsource=\regexp{$MAPLOOP}, match={XXX}, replace={Kibern. Vychisl. Tekh.}]
			\step[fieldsource=\regexp{$MAPLOOP}, match={XXX}, replace={Kibernetika}]
			\step[fieldsource=\regexp{$MAPLOOP}, match={XXX}, replace={Kristallografiya}]
			\step[fieldsource=\regexp{$MAPLOOP}, match={XXX}, replace={Kvantovaya Elektron. Mosk.}]
			\step[fieldsource=\regexp{$MAPLOOP}, match={XXX}, replace={Kybernetes}]
			\step[fieldsource=\regexp{$MAPLOOP}, match={XXX}, replace={Kybernetika}]
			\step[fieldsource=\regexp{$MAPLOOP}, match={XXX}, replace={Med. Tek.}]
			\step[fieldsource=\regexp{$MAPLOOP}, match={XXX}, replace={Mekh. Avtom. Proizvod.}]
			\step[fieldsource=\regexp{$MAPLOOP}, match={XXX}, replace={Meres Autom.}]
			\step[fieldsource=\regexp{$MAPLOOP}, match={XXX}, replace={Mesures}]
			\step[fieldsource=\regexp{$MAPLOOP}, match={XXX}, replace={Metallofizika}]
			\step[fieldsource=\regexp{$MAPLOOP}, match={XXX}, replace={Metalloved. Term. Obrab. Met.}]
			\step[fieldsource=\regexp{$MAPLOOP}, match={XXX}, replace={Meteorol. Gidrol.}]
			\step[fieldsource=\regexp{$MAPLOOP}, match={XXX}, replace={Meteorol. Rundsch.}]
			\step[fieldsource=\regexp{$MAPLOOP}, match={XXX}, replace={Metrol. Apl.}]
			\step[fieldsource=\regexp{$MAPLOOP}, match={XXX}, replace={Metrologia}]
			\step[fieldsource=\regexp{$MAPLOOP}, match={XXX}, replace={Medel. Simul.}]
			\step[fieldsource=\regexp{$MAPLOOP}, match={XXX}, replace={Nachr. Dok.}]
			\step[fieldsource=\regexp{$MAPLOOP}, match={XXX}, replace={Nachr.tech. Elektron.}]
			\step[fieldsource=\regexp{$MAPLOOP}, match={XXX}, replace={Naturwissentchaften}]
			\step[fieldsource=\regexp{$MAPLOOP}, match={XXX}, replace={Neue Tech.}]
			\step[fieldsource=\regexp{$MAPLOOP}, match={XXX}, replace={Neue Tech. Buero}]
			\step[fieldsource=\regexp{$MAPLOOP}, match={XXX}, replace={Nukleonika}]
			\step[fieldsource=\regexp{$MAPLOOP}, match={XXX}, replace={Numer. Math.}]
			\step[fieldsource=\regexp{$MAPLOOP}, match={XXX}, replace={Nuovo Cimento A}]
			\step[fieldsource=\regexp{$MAPLOOP}, match={XXX}, replace={Nuovo Cimento B}]
			\step[fieldsource=\regexp{$MAPLOOP}, match={XXX}, replace={Nouvo Cimento C}]
			\step[fieldsource=\regexp{$MAPLOOP}, match={XXX}, replace={Nuovo Cimento D}]
			\step[fieldsource=\regexp{$MAPLOOP}, match={XXX}, replace={Okeanologiya}]
			\step[fieldsource=\regexp{$MAPLOOP}, match={XXX}, replace={Opt. Spektrosk.}]
			\step[fieldsource=\regexp{$MAPLOOP}, match={XXX}, replace={Opt.-Mekh. Prom.}]
			\step[fieldsource=\regexp{$MAPLOOP}, match={XXX}, replace={Optik}]
			\step[fieldsource=\regexp{$MAPLOOP}, match={XXX}, replace={Photogrammetria}]
			\step[fieldsource=\regexp{$MAPLOOP}, match={XXX}, replace={Photonics Spectra}]
			\step[fieldsource=\regexp{$MAPLOOP}, match={XXX}, replace={Pis’ma Astron. Zh. }]
			\step[fieldsource=\regexp{$MAPLOOP}, match={XXX}, replace={Pis’ma Zh. Eksp. Teor. Fiz. }]
			\step[fieldsource=\regexp{$MAPLOOP}, match={XXX}, replace={Pis’ma Zh. Tekh. Fiz. }]
			\step[fieldsource=\regexp{$MAPLOOP}, match={XXX}, replace={Poverkhn., Fiz. Khim. Mekh.}]
			\step[fieldsource=\regexp{$MAPLOOP}, match={XXX}, replace={Pr. Inst. Elektrotech.}]
			\step[fieldsource=\regexp{$MAPLOOP}, match={XXX}, replace={Prib. Sist. Upr.}]
			\step[fieldsource=\regexp{$MAPLOOP}, match={XXX}, replace={Prib. Tekh. Eksp.}]
			\step[fieldsource=\regexp{$MAPLOOP}, match={XXX}, replace={Prikl. Mat. Mekh.}]
			\step[fieldsource=\regexp{$MAPLOOP}, match={XXX}, replace={Prikl. Mekh.}]
			\step[fieldsource=\regexp{$MAPLOOP}, match={XXX}, replace={Probl. Kibern.}]
			\step[fieldsource=\regexp{$MAPLOOP}, match={XXX}, replace={Probl. Peredachi Inf.}]
			\step[fieldsource=\regexp{$MAPLOOP}, match={XXX}, replace={Proc. K. Ned. Akad. Wet. B, Palaeontol.}]
			\step[fieldsource=\regexp{$MAPLOOP}, match={XXX}, replace={Anthropol. }]
			\step[fieldsource=\regexp{$MAPLOOP}, match={XXX}, replace={Programmirovanie}]
			\step[fieldsource=\regexp{$MAPLOOP}, match={XXX}, replace={Prz. Elektrotech.}]
			\step[fieldsource=\regexp{$MAPLOOP}, match={XXX}, replace={Prz. Telekomun.}]
			\step[fieldsource=\regexp{$MAPLOOP}, match={XXX}, replace={PT/Elektrotech. Elektron.}]
			\step[fieldsource=\regexp{$MAPLOOP}, match={XXX}, replace={Radio Fernsehen Elektron.}]
			\step[fieldsource=\regexp{$MAPLOOP}, match={XXX}, replace={Radiotekh. Elektron.}]
			\step[fieldsource=\regexp{$MAPLOOP}, match={XXX}, replace={Radiotekhnika Mosk.}]
			\step[fieldsource=\regexp{$MAPLOOP}, match={XXX}, replace={Rev. Acad. Cienc. Zaragoza}]
			\step[fieldsource=\regexp{$MAPLOOP}, match={XXX}, replace={Rev. Electrotec. (Argentina)}]
			\step[fieldsource=\regexp{$MAPLOOP}, match={XXX}, replace={Rev. Electrotec. (Spain)}]
			\step[fieldsource=\regexp{$MAPLOOP}, match={XXX}, replace={Rev. Energ.}]
			\step[fieldsource=\regexp{$MAPLOOP}, match={XXX}, replace={Rev. Esp. Electron. Rev. Geofis.}]
			\step[fieldsource=\regexp{$MAPLOOP}, match={XXX}, replace={Ric. Autom.}]
			\step[fieldsource=\regexp{$MAPLOOP}, match={XXX}, replace={Ric. Spettrosc.}]
			\step[fieldsource=\regexp{$MAPLOOP}, match={XXX}, replace={Robotersysteme}]
			\step[fieldsource=\regexp{$MAPLOOP}, match={XXX}, replace={Rozpr. Electrotech.}]
			\step[fieldsource=\regexp{$MAPLOOP}, match={XXX}, replace={Sadhana}]
			\step[fieldsource=\regexp{$MAPLOOP}, match={XXX}, replace={Schweiz. Tech. Z.}]
			\step[fieldsource=\regexp{$MAPLOOP}, match={XXX}, replace={Scientia}]
			\step[fieldsource=\regexp{$MAPLOOP}, match={XXX}, replace={Siemens Forsch. Entwickl. Ber.}]
			\step[fieldsource=\regexp{$MAPLOOP}, match={XXX}, replace={Sist. Autom.}]
			\step[fieldsource=\regexp{$MAPLOOP}, match={XXX}, replace={Sitzungsber. Oester. Akad. Wiss. Math.-}]
			\step[fieldsource=\regexp{$MAPLOOP}, match={XXX}, replace={Naturwiss. Kl. Abt. II (Austria)}]
			\step[fieldsource=\regexp{$MAPLOOP}, match={XXX}, replace={Spectrochim. Acta A, Mol. Spectrosc.}]
			\step[fieldsource=\regexp{$MAPLOOP}, match={XXX}, replace={Spectrochim. Acta B, At. Spectrosc.}]
			\step[fieldsource=\regexp{$MAPLOOP}, match={XXX}, replace={Sprache Datenverarb.}]
			\step[fieldsource=\regexp{$MAPLOOP}, match={XXX}, replace={Stanki Instrum.}]
			\step[fieldsource=\regexp{$MAPLOOP}, match={XXX}, replace={Steklo Keram.}]
			\step[fieldsource=\regexp{$MAPLOOP}, match={XXX}, replace={Svetotekhnika  TE Int.}]
			\step[fieldsource=\regexp{$MAPLOOP}, match={XXX}, replace={Tech. Bull. Vevey}]
			\step[fieldsource=\regexp{$MAPLOOP}, match={XXX}, replace={Tech. Mitt. Krupp (Engl. Ed.)}]
			\step[fieldsource=\regexp{$MAPLOOP}, match={XXX}, replace={Tech. Mitt. PTT}]
			\step[fieldsource=\regexp{$MAPLOOP}, match={XXX}, replace={Tech. Mitt. RFZ}]
			\step[fieldsource=\regexp{$MAPLOOP}, match={XXX}, replace={Technica}]
			\step[fieldsource=\regexp{$MAPLOOP}, match={XXX}, replace={Tecnica}]
			\step[fieldsource=\regexp{$MAPLOOP}, match={XXX}, replace={Teh. Fiz.}]
			\step[fieldsource=\regexp{$MAPLOOP}, match={XXX}, replace={Tehnika}]
			\step[fieldsource=\regexp{$MAPLOOP}, match={XXX}, replace={Tekh. Elektrodin.}]
			\step[fieldsource=\regexp{$MAPLOOP}, match={XXX}, replace={Tekh. Kibern.}]
			\step[fieldsource=\regexp{$MAPLOOP}, match={XXX}, replace={Tekh. Kino Telev.}]
			\step[fieldsource=\regexp{$MAPLOOP}, match={XXX}, replace={Tekh. Misul}]
			\step[fieldsource=\regexp{$MAPLOOP}, match={XXX}, replace={Telekomunikacije}]
			\step[fieldsource=\regexp{$MAPLOOP}, match={XXX}, replace={Telektronikk}]
			\step[fieldsource=\regexp{$MAPLOOP}, match={XXX}, replace={Teleteknik}]
			\step[fieldsource=\regexp{$MAPLOOP}, match={XXX}, replace={Teor. Mat. Fiz.}]
			\step[fieldsource=\regexp{$MAPLOOP}, match={XXX}, replace={Teploeoergetika}]
			\step[fieldsource=\regexp{$MAPLOOP}, match={XXX}, replace={Teplofiz. Vvs. Temp.}]
			\step[fieldsource=\regexp{$MAPLOOP}, match={XXX}, replace={Tidskr. Dok.}]
			\step[fieldsource=\regexp{$MAPLOOP}, match={XXX}, replace={TN Nachr.}]
			\step[fieldsource=\regexp{$MAPLOOP}, match={XXX}, replace={Toute Electron.}]
			\step[fieldsource=\regexp{$MAPLOOP}, match={XXX}, replace={Tr. Inst. Teor. Astron.}]
			\step[fieldsource=\regexp{$MAPLOOP}, match={XXX}, replace={Ukr. Fiz. Zh.}]
			\step[fieldsource=\regexp{$MAPLOOP}, match={XXX}, replace={Usp. Fiz. Nauk}]
			\step[fieldsource=\regexp{$MAPLOOP}, match={XXX}, replace={Vak.-Tech.}]
			\step[fieldsource=\regexp{$MAPLOOP}, match={XXX}, replace={VDE Fachiber.}]
			\step[fieldsource=\regexp{$MAPLOOP}, match={XXX}, replace={VDI Z.}]
			\step[fieldsource=\regexp{$MAPLOOP}, match={XXX}, replace={Vestn. Mashinostr.}]
			\step[fieldsource=\regexp{$MAPLOOP}, match={XXX}, replace={Vestn. Mosk. Univ. 15, Vychisl. Mat. Kibern.}]
			\step[fieldsource=\regexp{$MAPLOOP}, match={XXX}, replace={Vestn. Mosk. Univ. 3, Fiz. Astron.}]
			\step[fieldsource=\regexp{$MAPLOOP}, match={XXX}, replace={Vesti Akad. Navuk BSSR Ser. Fiz. Energ. Navuk}]
			\step[fieldsource=\regexp{$MAPLOOP}, match={XXX}, replace={VGB Kraftwerkstech. (Ger. Ed.)}]
			\step[fieldsource=\regexp{$MAPLOOP}, match={XXX}, replace={Vide Couches Minces}]
			\step[fieldsource=\regexp{$MAPLOOP}, match={XXX}, replace={Vistas Astron.}]
			\step[fieldsource=\regexp{$MAPLOOP}, match={XXX}, replace={Vopr. At. Nauki Tekh. Ser., Fiz. Radiats.}]
			\step[fieldsource=\regexp{$MAPLOOP}, match={XXX}, replace={Povrezhdenii Radiats. Materialoved.}]
			\step[fieldsource=\regexp{$MAPLOOP}, match={XXX}, replace={Vopr. At. Nauki Tekh. Ser., Obshch. Yad. Fiz.}]
			\step[fieldsource=\regexp{$MAPLOOP}, match={XXX}, replace={Vuoto Sci. Tecnol.}]
			\step[fieldsource=\regexp{$MAPLOOP}, match={XXX}, replace={Wiss. Z. Friedrich-Schiller-Univ. Jena}]
			\step[fieldsource=\regexp{$MAPLOOP}, match={XXX}, replace={Nat.wiss. Reihe}]
			\step[fieldsource=\regexp{$MAPLOOP}, match={XXX}, replace={Wiss. Z. Karl-Marx-Univ. Leipz. Math.-}]
			\step[fieldsource=\regexp{$MAPLOOP}, match={XXX}, replace={Nat.wiss. Reihe}]
			\step[fieldsource=\regexp{$MAPLOOP}, match={XXX}, replace={Wiss. Z. Tech. Hochsch. Ilmenau}]
			\step[fieldsource=\regexp{$MAPLOOP}, match={XXX}, replace={Wiss. Z. Tech. Univ. Dresd.}]
			\step[fieldsource=\regexp{$MAPLOOP}, match={XXX}, replace={Wiss. Z. Tech. Univ. Karl-Marx-Stadt}]
			\step[fieldsource=\regexp{$MAPLOOP}, match={XXX}, replace={Yad. Fiz.}]
			\step[fieldsource=\regexp{$MAPLOOP}, match={XXX}, replace={Z. Angew. Math. Mech.}]
			\step[fieldsource=\regexp{$MAPLOOP}, match={XXX}, replace={Z. Angew. Math. Phys.}]
			\step[fieldsource=\regexp{$MAPLOOP}, match={XXX}, replace={Z. Met.kd.}]
			\step[fieldsource=\regexp{$MAPLOOP}, match={XXX}, replace={Z. Nat. Forsch. A, Phys. Phys. Chem. Kosmophys.}]
			\step[fieldsource=\regexp{$MAPLOOP}, match={XXX}, replace={Z. Oper. Res. A, Theor.}]
			\step[fieldsource=\regexp{$MAPLOOP}, match={XXX}, replace={Z. Oper. Res. B, Prax.}]
			\step[fieldsource=\regexp{$MAPLOOP}, match={XXX}, replace={Z. Phys.}]
			\step[fieldsource=\regexp{$MAPLOOP}, match={XXX}, replace={Z. Phys. A, At. Nuclei}]
			\step[fieldsource=\regexp{$MAPLOOP}, match={XXX}, replace={Z. Phys. B, Condens. Matter}]
			\step[fieldsource=\regexp{$MAPLOOP}, match={XXX}, replace={Z. Phys. C, Part Fields}]
			\step[fieldsource=\regexp{$MAPLOOP}, match={XXX}, replace={Z. Phys. Chem. Neue Folge}]
			\step[fieldsource=\regexp{$MAPLOOP}, match={XXX}, replace={Z. Phys. Chem., Leipz.}]
			\step[fieldsource=\regexp{$MAPLOOP}, match={XXX}, replace={Z. Phys. D, At. Mol. Clusters}]
			\step[fieldsource=\regexp{$MAPLOOP}, match={XXX}, replace={Zavod. Lab.}]
			\step[fieldsource=\regexp{$MAPLOOP}, match={XXX}, replace={Zh. Eksp. Teor. Fiz.}]
			\step[fieldsource=\regexp{$MAPLOOP}, match={XXX}, replace={Zh. Fiz. Khim.}]
			\step[fieldsource=\regexp{$MAPLOOP}, match={XXX}, replace={Zh. Prikl. Mekh. Tekh. Fiz.}]
			\step[fieldsource=\regexp{$MAPLOOP}, match={XXX}, replace={Zh. Prikl. Spektrosk.}]
			\step[fieldsource=\regexp{$MAPLOOP}, match={XXX}, replace={Zh. Tekh. Fiz.}]
			\step[fieldsource=\regexp{$MAPLOOP}, match={XXX}, replace={Zh. Vychisl. Mat. Mat. Fiz.}]
			\step[fieldsource=\regexp{$MAPLOOP}, match={XXX}, replace={Zisin, J. Seismol. Soc. Jpn.}] 
			\step[fieldsource=\regexp{$MAPLOOP}, match={Production}, replace={Prod.}]
			\step[fieldsource=\regexp{$MAPLOOP}, match={Reliability}, replace={Rel.}]
			\step[fieldsource=\regexp{$MAPLOOP}, match={Report}, replace={Rep.}]
			\step[fieldsource=\regexp{$MAPLOOP}, match={Semiconductor}, replace={Semicond.}]
			\step[fieldsource=\regexp{$MAPLOOP}, match={Research}, replace={Res.}]
			\step[fieldsource=\regexp{$MAPLOOP}, match={Sensing}, replace={Sens.}]
			\step[fieldsource=\regexp{$MAPLOOP}, match={Resonance}, replace={Reson.}]
			\step[fieldsource=\regexp{$MAPLOOP}, match={Series}, replace={Ser.}]
			\step[fieldsource=\regexp{$MAPLOOP}, match={Resources}, replace={Resour.}]
			\step[fieldsource=\regexp{$MAPLOOP}, match={Simulation}, replace={Simul.}]
			\step[fieldsource=\regexp{$MAPLOOP}, match={Reviews}, replace={Rev.}]
			\step[fieldsource=\regexp{$MAPLOOP}, match={Review}, replace={Rev.}]
			\step[fieldsource=\regexp{$MAPLOOP}, match={Singapore}, replace={Singap.}]
			\step[fieldsource=\regexp{$MAPLOOP}, match={Robotics}, replace={Robot.}]
			\step[fieldsource=\regexp{$MAPLOOP}, match={Sistema}, replace={Sist.}]
			\step[fieldsource=\regexp{$MAPLOOP}, match={Royal}, replace={Roy.}]
			\step[fieldsource=\regexp{$MAPLOOP}, match={Society}, replace={Soc.}]
			\step[fieldsource=\regexp{$MAPLOOP}, match={Safety}, replace={Saf.}]
			\step[fieldsource=\regexp{$MAPLOOP}, match={Sociological}, replace={Sociol.}]
			\step[fieldsource=\regexp{$MAPLOOP}, match={Satellite}, replace={Satell.}]
			\step[fieldsource=\regexp{$MAPLOOP}, match={Software}, replace={Softw.}]
			\step[fieldsource=\regexp{$MAPLOOP}, match={Scandinavian}, replace={Scand.}]
			\step[fieldsource=\regexp{$MAPLOOP}, match={Solar}, replace={Sol.}]
			\step[fieldsource=\regexp{$MAPLOOP}, match={Sciences}, replace={Sci.}]
			\step[fieldsource=\regexp{$MAPLOOP}, match={Science}, replace={Sci.}]
			\step[fieldsource=\regexp{$MAPLOOP}, match={Soviet}, replace={Sov.}]
			\step[fieldsource=\regexp{$MAPLOOP}, match={Section}, replace={Sect.}]
			\step[fieldsource=\regexp{$MAPLOOP}, match={Spectroscopy}, replace={Spectrosc.}]
			\step[fieldsource=\regexp{$MAPLOOP}, match={Security}, replace={Secur.}]
			\step[fieldsource=\regexp{$MAPLOOP}, match={Spectrum}, replace={Spectr.}]
			\step[fieldsource=\regexp{$MAPLOOP}, match={Seismology}, replace={Seismol.}]
			\step[fieldsource=\regexp{$MAPLOOP}, match={Speculations}, replace={Specul.}]
			\step[fieldsource=\regexp{$MAPLOOP}, match={Selected}, replace={Sel.}]
			\step[fieldsource=\regexp{$MAPLOOP}, match={Statistics}, replace={Statist.}]
			\step[fieldsource=\regexp{$MAPLOOP}, match={Structures}, replace={Struct.}]
			\step[fieldsource=\regexp{$MAPLOOP}, match={Structure}, replace={Struct.}]
			\step[fieldsource=\regexp{$MAPLOOP}, match={Terrestrial}, replace={Terr.}]
			\step[fieldsource=\regexp{$MAPLOOP}, match={Studies}, replace={Stud.}]
			\step[fieldsource=\regexp{$MAPLOOP}, match={Theoretical}, replace={Theor.}]
			\step[fieldsource=\regexp{$MAPLOOP}, match={Superconductivity}, replace={Supercond.}]
			\step[fieldsource=\regexp{$MAPLOOP}, match={Transactions}, replace={Trans.}]
			\step[fieldsource=\regexp{$MAPLOOP}, match={Supplement}, replace={Suppl.}]
			\step[fieldsource=\regexp{$MAPLOOP}, match={Translation}, replace={Transl.}]
			\step[fieldsource=\regexp{$MAPLOOP}, match={Surface}, replace={Surf.}]
			\step[fieldsource=\regexp{$MAPLOOP}, match={Transmission}, replace={Transmiss.}]
			\step[fieldsource=\regexp{$MAPLOOP}, match={Survey}, replace={Surv.}]
			\step[fieldsource=\regexp{$MAPLOOP}, match={Transportation}, replace={Transp.}]
			\step[fieldsource=\regexp{$MAPLOOP}, match={Sustainable}, replace={Sustain.}]
			\step[fieldsource=\regexp{$MAPLOOP}, match={Tutorials}, replace={Tut.}]
			\step[fieldsource=\regexp{$MAPLOOP}, match={Symposium}, replace={Symp.}]
			\step[fieldsource=\regexp{$MAPLOOP}, match={Ultrasonic}, replace={Ultrason.}]
			\step[fieldsource=\regexp{$MAPLOOP}, match={Systems}, replace={Syst.}]
			\step[fieldsource=\regexp{$MAPLOOP}, match={System}, replace={Syst.}]
			\step[fieldsource=\regexp{$MAPLOOP}, match={University}, replace={Univ.}]
			\step[fieldsource=\regexp{$MAPLOOP}, match={\detokenize{Universität}}, replace={Univ.}]
			\step[fieldsource=\regexp{$MAPLOOP}, match={\detokenize{Université}}, replace={Univ.}]
			\step[fieldsource=\regexp{$MAPLOOP}, match={Technical}, replace={Tech.}]
			\step[fieldsource=\regexp{$MAPLOOP}, match={Technische}, replace={Tech.}]
			\step[fieldsource=\regexp{$MAPLOOP}, match={Vacuum}, replace={Vac.}]
			\step[fieldsource=\regexp{$MAPLOOP}, match={Techniques}, replace={Techn.}]
			\step[fieldsource=\regexp{$MAPLOOP}, match={Vehicles}, replace={Veh.}]
			\step[fieldsource=\regexp{$MAPLOOP}, match={Vehicle}, replace={Veh.}]
			\step[fieldsource=\regexp{$MAPLOOP}, match={Vehicular}, replace={Veh.}]
			\step[fieldsource=\regexp{$MAPLOOP}, match={Technology}, replace={Technol.}]
			\step[fieldsource=\regexp{$MAPLOOP}, match={Technological}, replace={Technol.}]
			\step[fieldsource=\regexp{$MAPLOOP}, match={Vibration}, replace={Vib.}]
			\step[fieldsource=\regexp{$MAPLOOP}, match={Telecommunications}, replace={Telecommun.}]
			\step[fieldsource=\regexp{$MAPLOOP}, match={Visual}, replace={Vis.}]
			\step[fieldsource=\regexp{$MAPLOOP}, match={Television}, replace={Telev.}]
			\step[fieldsource=\regexp{$MAPLOOP}, match={Welding}, replace={Weld.}]
			\step[fieldsource=\regexp{$MAPLOOP}, match={Temperature}, replace={Temp.}]
			\step[fieldsource=\regexp{$MAPLOOP}, match={Working}, replace={Work.}]
			\step[fieldsource=\regexp{$MAPLOOP}, match={Learning}, replace={Learn.}]
			\step[fieldsource=\regexp{$MAPLOOP}, match={Measurement}, replace={Meas.}]
			\step[fieldsource=\regexp{$MAPLOOP}, match={Letters}, replace={Lett.}]
			\step[fieldsource=\regexp{$MAPLOOP}, match={Letter}, replace={Lett.}]
			\step[fieldsource=\regexp{$MAPLOOP}, match={Mechanical}, replace={Mech.}]
			\step[fieldsource=\regexp{$MAPLOOP}, match={Mechanics}, replace={Mech.}]
			\step[fieldsource=\regexp{$MAPLOOP}, match={Mechanic}, replace={Mech.}]
			\step[fieldsource=\regexp{$MAPLOOP}, match={Lightwave}, replace={Lightw.}]
			\step[fieldsource=\regexp{$MAPLOOP}, match={Medical}, replace={Med.}]
			\step[fieldsource=\regexp{$MAPLOOP}, match={Logic, Logical}, replace={Log.}]
			\step[fieldsource=\regexp{$MAPLOOP}, match={Metals}, replace={Met.}]
			\step[fieldsource=\regexp{$MAPLOOP}, match={Luminescence}, replace={Lumin.}]
			\step[fieldsource=\regexp{$MAPLOOP}, match={Metallurgy}, replace={Metall.}]
			\step[fieldsource=\regexp{$MAPLOOP}, match={Machines}, replace={Mach.}]
			\step[fieldsource=\regexp{$MAPLOOP}, match={Machine}, replace={Mach.}]
			\step[fieldsource=\regexp{$MAPLOOP}, match={Meteorology}, replace={Meteorol.}]
			\step[fieldsource=\regexp{$MAPLOOP}, match={Magazine}, replace={Mag.}]
			\step[fieldsource=\regexp{$MAPLOOP}, match={Metropolitan}, replace={Metrop.}]
			\step[fieldsource=\regexp{$MAPLOOP}, match={Magnetics}, replace={Magn.}]
			\step[fieldsource=\regexp{$MAPLOOP}, match={Mexican, Mexico}, replace={Mex.}]
			\step[fieldsource=\regexp{$MAPLOOP}, match={Management}, replace={Manage.}]
			\step[fieldsource=\regexp{$MAPLOOP}, match={Microelectromechanical}, replace={Microelectromech.}]
			\step[fieldsource=\regexp{$MAPLOOP}, match={Managing}, replace={Manag.}]
			\step[fieldsource=\regexp{$MAPLOOP}, match={Microgravity}, replace={Microgr.}]
			\step[fieldsource=\regexp{$MAPLOOP}, match={Manufacturing}, replace={Manuf.}]
			\step[fieldsource=\regexp{$MAPLOOP}, match={Microscopy}, replace={Microsc.}]
			\step[fieldsource=\regexp{$MAPLOOP}, match={Marine}, replace={Mar.}]
			\step[fieldsource=\regexp{$MAPLOOP}, match={Microwaves}, replace={Microw.}]
			\step[fieldsource=\regexp{$MAPLOOP}, match={Microwave}, replace={Microw.}]
			\step[fieldsource=\regexp{$MAPLOOP}, match={Materials}, replace={Mater.}]
			\step[fieldsource=\regexp{$MAPLOOP}, match={Material}, replace={Mater.}]
			\step[fieldsource=\regexp{$MAPLOOP}, match={Military}, replace={Mil.}]
			\step[fieldsource=\regexp{$MAPLOOP}, match={Modeling}, replace={Model.}]
			\step[fieldsource=\regexp{$MAPLOOP}, match={Modelling}, replace={Model.}]
			\step[fieldsource=\regexp{$MAPLOOP}, match={Oceanic}, replace={Ocean.}]
			\step[fieldsource=\regexp{$MAPLOOP}, match={Molecular}, replace={Mol.}]
			\step[fieldsource=\regexp{$MAPLOOP}, match={Oceanography}, replace={Oceanogr.}]
			\step[fieldsource=\regexp{$MAPLOOP}, match={Monitoring}, replace={Monit.}]
			\step[fieldsource=\regexp{$MAPLOOP}, match={Occupation}, replace={Occupat.}]
			\step[fieldsource=\regexp{$MAPLOOP}, match={Multiphysics}, replace={Multiphys.}]
			\step[fieldsource=\regexp{$MAPLOOP}, match={Operational}, replace={Oper.}]
			\step[fieldsource=\regexp{$MAPLOOP}, match={Nanobioscience}, replace={Nanobiosci.}]
			\step[fieldsource=\regexp{$MAPLOOP}, match={Optical}, replace={Opt.}]
			\step[fieldsource=\regexp{$MAPLOOP}, match={Nanotechnology}, replace={Nanotechnol.}]
			\step[fieldsource=\regexp{$MAPLOOP}, match={Optics}, replace={Opt.}]
			\step[fieldsource=\regexp{$MAPLOOP}, match={National}, replace={Nat.}]
			\step[fieldsource=\regexp{$MAPLOOP}, match={Optimization}, replace={Optim.}]
			\step[fieldsource=\regexp{$MAPLOOP}, match={Naval}, replace={Nav.}]
			\step[fieldsource=\regexp{$MAPLOOP}, match={Organization}, replace={Org.}]
			\step[fieldsource=\regexp{$MAPLOOP}, match={Networking}, replace={Netw.}]
			\step[fieldsource=\regexp{$MAPLOOP}, match={Networked}, replace={Netw.}]
			\step[fieldsource=\regexp{$MAPLOOP}, match={Network}, replace={Netw.}]
			\step[fieldsource=\regexp{$MAPLOOP}, match={Packaging}, replace={Packag.}]
			\step[fieldsource=\regexp{$MAPLOOP}, match={Newsletter}, replace={Newslett.}]
			\step[fieldsource=\regexp{$MAPLOOP}, match={Particle}, replace={Part.}]
			\step[fieldsource=\regexp{$MAPLOOP}, match={Nondestructive}, replace={Nondestruct.}]
			\step[fieldsource=\regexp{$MAPLOOP}, match={Patent}, replace={Pat.}]
			\step[fieldsource=\regexp{$MAPLOOP}, match={Nuclear}, replace={Nucl.}]
			\step[fieldsource=\regexp{$MAPLOOP}, match={Performance}, replace={Perform.}]
			\step[fieldsource=\regexp{$MAPLOOP}, match={Numerical}, replace={Numer.}]
			\step[fieldsource=\regexp{$MAPLOOP}, match={Personal}, replace={Pers.}]
			\step[fieldsource=\regexp{$MAPLOOP}, match={Observations}, replace={Observ.}]
			\step[fieldsource=\regexp{$MAPLOOP}, match={Philosophical}, replace={Philos.}]
			\step[fieldsource=\regexp{$MAPLOOP}, match={Photonics}, replace={Photon.}]
			\step[fieldsource=\regexp{$MAPLOOP}, match={Productivity}, replace={Productiv.}]
			\step[fieldsource=\regexp{$MAPLOOP}, match={Photovoltaics}, replace={Photovolt.}]
			\step[fieldsource=\regexp{$MAPLOOP}, match={Programming}, replace={Program.}]
			\step[fieldsource=\regexp{$MAPLOOP}, match={Physics}, replace={Phys.}]
			\step[fieldsource=\regexp{$MAPLOOP}, match={Progress}, replace={Prog.}]
			\step[fieldsource=\regexp{$MAPLOOP}, match={Physiology}, replace={Physiol.}]
			\step[fieldsource=\regexp{$MAPLOOP}, match={Propagation}, replace={Propag.}]
			\step[fieldsource=\regexp{$MAPLOOP}, match={Planetary}, replace={Planet.}]
			\step[fieldsource=\regexp{$MAPLOOP}, match={Psychology}, replace={Psychol.}]
			\step[fieldsource=\regexp{$MAPLOOP}, match={Pneumatics}, replace={Pneum.}]
			\step[fieldsource=\regexp{$MAPLOOP}, match={Quality}, replace={Qual.}]
			\step[fieldsource=\regexp{$MAPLOOP}, match={Pollution}, replace={Pollut.}]
			\step[fieldsource=\regexp{$MAPLOOP}, match={Quarterly}, replace={Quart.}]
			\step[fieldsource=\regexp{$MAPLOOP}, match={Polymer}, replace={Polym.}]
			\step[fieldsource=\regexp{$MAPLOOP}, match={Radiation}, replace={Radiat.}]
			\step[fieldsource=\regexp{$MAPLOOP}, match={Polytechnic}, replace={Polytech.}]
			\step[fieldsource=\regexp{$MAPLOOP}, match={Radiology}, replace={Radiol.}]
			\step[fieldsource=\regexp{$MAPLOOP}, match={Practice}, replace={Pract.}]
			\step[fieldsource=\regexp{$MAPLOOP}, match={Reactor}, replace={React.}]
			\step[fieldsource=\regexp{$MAPLOOP}, match={Precision}, replace={Precis.}]
			\step[fieldsource=\regexp{$MAPLOOP}, match={Receivers}, replace={Receiv.}]
			\step[fieldsource=\regexp{$MAPLOOP}, match={Principles}, replace={Princ.}]
			\step[fieldsource=\regexp{$MAPLOOP}, match={Recognition}, replace={Recognit.}]
			\step[fieldsource=\regexp{$MAPLOOP}, match={Proceedings}, replace={Proc.}]
			\step[fieldsource=\regexp{$MAPLOOP}, match={Record}, replace={Rec.}]
			\step[fieldsource=\regexp{$MAPLOOP}, match={Processing}, replace={Process.}]
			\step[fieldsource=\regexp{$MAPLOOP}, match={Rehabilitation}, replace={Rehabil.}]
			\step[fieldsource=\regexp{$MAPLOOP}, match={Conversion}, replace={Convers.}]
			\step[fieldsource=\regexp{$MAPLOOP}, match={Digital}, replace={Digit.}]
			\step[fieldsource=\regexp{$MAPLOOP}, match={Convention}, replace={Conv.}]
			\step[fieldsource=\regexp{$MAPLOOP}, match={Disclosure}, replace={Discl.}]
			\step[fieldsource=\regexp{$MAPLOOP}, match={Correspondence}, replace={Corresp.}]
			\step[fieldsource=\regexp{$MAPLOOP}, match={Discussions}, replace={Discuss.}]
			\step[fieldsource=\regexp{$MAPLOOP}, match={Critical}, replace={Crit.}]
			\step[fieldsource=\regexp{$MAPLOOP}, match={Dissertations}, replace={Diss.}]
			\step[fieldsource=\regexp{$MAPLOOP}, match={Crystal}, replace={Cryst.}]
			\step[fieldsource=\regexp{$MAPLOOP}, match={Distributed}, replace={Distrib.}]
			\step[fieldsource=\regexp{$MAPLOOP}, match={Crystallography}, replace={Crystallogr.}]
			\step[fieldsource=\regexp{$MAPLOOP}, match={Dynamics}, replace={Dyn.}]
			\step[fieldsource=\regexp{$MAPLOOP}, match={Cybernetics}, replace={Cybern.}]
			\step[fieldsource=\regexp{$MAPLOOP}, match={Earthquake}, replace={Earthq.}]
			\step[fieldsource=\regexp{$MAPLOOP}, match={Decision}, replace={Decis.}]
			\step[fieldsource=\regexp{$MAPLOOP}, match={Economics}, replace={Econ.}]
			\step[fieldsource=\regexp{$MAPLOOP}, match={Economic}, replace={Econ.}]
			\step[fieldsource=\regexp{$MAPLOOP}, match={Economical}, replace={Econ.}]
			\step[fieldsource=\regexp{$MAPLOOP}, match={Edition}, replace={Ed.}]
			\step[fieldsource=\regexp{$MAPLOOP}, match={Evolutionary}, replace={Evol.}]
			\step[fieldsource=\regexp{$MAPLOOP}, match={Education}, replace={Educ.}]
			\step[fieldsource=\regexp{$MAPLOOP}, match={Exhibition}, replace={Exhib.}]
			\step[fieldsource=\regexp{$MAPLOOP}, match={Electrical}, replace={Elect.}]
			\step[fieldsource=\regexp{$MAPLOOP}, match={Electric}, replace={Elect.}]
			\step[fieldsource=\regexp{$MAPLOOP}, match={Experimental}, replace={Exp.}]
			\step[fieldsource=\regexp{$MAPLOOP}, match={Electrification}, replace={Electrific.}]
			\step[fieldsource=\regexp{$MAPLOOP}, match={Exploratory}, replace={Explor.}]
			\step[fieldsource=\regexp{$MAPLOOP}, match={Electromagnetic}, replace={Electromagn.}]
			\step[fieldsource=\regexp{$MAPLOOP}, match={Exposition}, replace={Expo.}]
			\step[fieldsource=\regexp{$MAPLOOP}, match={Electroacoustic}, replace={Electroacoust.}]
			\step[fieldsource=\regexp{$MAPLOOP}, match={Express}, replace={Express}]
			\step[fieldsource=\regexp{$MAPLOOP}, match={Electronics}, replace={Electron.}]
			\step[fieldsource=\regexp{$MAPLOOP}, match={Electronic}, replace={Electron.}]
			\step[fieldsource=\regexp{$MAPLOOP}, match={Fabrication}, replace={Fabr.}]
			\step[fieldsource=\regexp{$MAPLOOP}, match={Emerging}, replace={Emerg.}]
			\step[fieldsource=\regexp{$MAPLOOP}, match={Faculty}, replace={Fac.}]
			\step[fieldsource=\regexp{$MAPLOOP}, match={Engineering}, replace={Eng.}]
			\step[fieldsource=\regexp{$MAPLOOP}, match={Engineers}, replace={Eng.}]
			\step[fieldsource=\regexp{$MAPLOOP}, match={Engineer}, replace={Eng.}]
			\step[fieldsource=\regexp{$MAPLOOP}, match={Ferroelectrics}, replace={Ferroelect.}]
			\step[fieldsource=\regexp{$MAPLOOP}, match={Environment}, replace={Environ.}]
			\step[fieldsource=\regexp{$MAPLOOP}, match={Francais, French}, replace={Fr.}]
			\step[fieldsource=\regexp{$MAPLOOP}, match={Equations}, replace={Equ.}]
			\step[fieldsource=\regexp{$MAPLOOP}, match={Frequency}, replace={Freq.}]
			\step[fieldsource=\regexp{$MAPLOOP}, match={Equipment}, replace={Equip.}]
			\step[fieldsource=\regexp{$MAPLOOP}, match={Foundation}, replace={Found.}]
			\step[fieldsource=\regexp{$MAPLOOP}, match={Ergonomics}, replace={Ergonom.}]
			\step[fieldsource=\regexp{$MAPLOOP}, match={Fundamental}, replace={Fundam.}]
			\step[fieldsource=\regexp{$MAPLOOP}, match={European}, replace={Eur.}]
			\step[fieldsource=\regexp{$MAPLOOP}, match={Generation}, replace={Gener.}]
			\step[fieldsource=\regexp{$MAPLOOP}, match={Evaluation}, replace={Eval.}]
			\step[fieldsource=\regexp{$MAPLOOP}, match={Geology}, replace={Geol.}]
			\step[fieldsource=\regexp{$MAPLOOP}, match={Geophysics}, replace={Geophys.}]
			\step[fieldsource=\regexp{$MAPLOOP}, match={Innovations}, replace={Innov.}]
			\step[fieldsource=\regexp{$MAPLOOP}, match={Innovation}, replace={Innov.}]
			\step[fieldsource=\regexp{$MAPLOOP}, match={Geoscience}, replace={Geosci.}]
			\step[fieldsource=\regexp{$MAPLOOP}, match={Institute}, replace={Inst.}]
			\step[fieldsource=\regexp{$MAPLOOP}, match={Institution}, replace={Inst.}]
			\step[fieldsource=\regexp{$MAPLOOP}, match={Graphics}, replace={Graph.}]
			\step[fieldsource=\regexp{$MAPLOOP}, match={Instrument}, replace={Instrum.}]
			\step[fieldsource=\regexp{$MAPLOOP}, match={Guidance}, replace={Guid.}]
			\step[fieldsource=\regexp{$MAPLOOP}, match={Instrumentation}, replace={Instrum.}]
			\step[fieldsource=\regexp{$MAPLOOP}, match={Harmonics}, replace={Harmon.}]
			\step[fieldsource=\regexp{$MAPLOOP}, match={Harmonic}, replace={Harmon.}]
			\step[fieldsource=\regexp{$MAPLOOP}, match={Insulation}, replace={Insul.}]
			\step[fieldsource=\regexp{$MAPLOOP}, match={History}, replace={Hist.}]
			\step[fieldsource=\regexp{$MAPLOOP}, match={Integrated}, replace={Integr.}]
			\step[fieldsource=\regexp{$MAPLOOP}, match={Horizon}, replace={Horiz.}]
			\step[fieldsource=\regexp{$MAPLOOP}, match={Intelligence}, replace={Intell.}]
			\step[fieldsource=\regexp{$MAPLOOP}, match={Hungary}, replace={Hung.}]
			\step[fieldsource=\regexp{$MAPLOOP}, match={Hungarian}, replace={Hung.}]
			\step[fieldsource=\regexp{$MAPLOOP}, match={Intelligent}, replace={Intell.}]
			\step[fieldsource=\regexp{$MAPLOOP}, match={Hydraulics}, replace={Hydraul.}]
			\step[fieldsource=\regexp{$MAPLOOP}, match={Interactions}, replace={Interact.}]
			\step[fieldsource=\regexp{$MAPLOOP}, match={Hydrology}, replace={Hydrol.}]
			\step[fieldsource=\regexp{$MAPLOOP}, match={Internationales}, replace={Int.}]
			\step[fieldsource=\regexp{$MAPLOOP}, match={International}, replace={Int.}]
			\step[fieldsource=\regexp{$MAPLOOP}, match={Illuminating}, replace={Illum.}]
			\step[fieldsource=\regexp{$MAPLOOP}, match={Isotopes}, replace={Isot.}]
			\step[fieldsource=\regexp{$MAPLOOP}, match={Imaging}, replace={Imag.}]
			\step[fieldsource=\regexp{$MAPLOOP}, match={Israel}, replace={Isr.}]
			\step[fieldsource=\regexp{$MAPLOOP}, match={Industrial}, replace={Ind.}]
			\step[fieldsource=\regexp{$MAPLOOP}, match={Japan}, replace={Jpn.}]
			\step[fieldsource=\regexp{$MAPLOOP}, match={Information}, replace={Inf.}]
			\step[fieldsource=\regexp{$MAPLOOP}, match={Journal}, replace={J.}]
			\step[fieldsource=\regexp{$MAPLOOP}, match={Informatics}, replace={Inform.}]
			\step[fieldsource=\regexp{$MAPLOOP}, match={Knowledge}, replace={Knowl.}]
			\step[fieldsource=\regexp{$MAPLOOP}, match={Laboratory}, replace={Lab.}]
			\step[fieldsource=\regexp{$MAPLOOP}, match={Laboratories}, replace={Lab.}]
			\step[fieldsource=\regexp{$MAPLOOP}, match={Mathematical}, replace={Math.}]
			\step[fieldsource=\regexp{$MAPLOOP}, match={Language}, replace={Lang.}]
			\step[fieldsource=\regexp{$MAPLOOP}, match={Mathematics}, replace={Math.}]
			\step[fieldsource=\regexp{$MAPLOOP}, match={Abstracts}, replace={Abstr.}]
			\step[fieldsource=\regexp{$MAPLOOP}, match={Analysis}, replace={Anal.}]
			\step[fieldsource=\regexp{$MAPLOOP}, match={Academy}, replace={Acad.}]
			\step[fieldsource=\regexp{$MAPLOOP}, match={Annals}, replace={Ann.}]
			\step[fieldsource=\regexp{$MAPLOOP}, match={Accelerator}, replace={Accel.}]
			\step[fieldsource=\regexp{$MAPLOOP}, match={Annual}, replace={Annu.}]
			\step[fieldsource=\regexp{$MAPLOOP}, match={Acoustics}, replace={Acoust.}]
			\step[fieldsource=\regexp{$MAPLOOP}, match={Apparatus}, replace={App.}]
			\step[fieldsource=\regexp{$MAPLOOP}, match={Active}, replace={Act.}]
			\step[fieldsource=\regexp{$MAPLOOP}, match={Applications}, replace={Appl.}]
			\step[fieldsource=\regexp{$MAPLOOP}, match={Administration}, replace={Admin.}]
			\step[fieldsource=\regexp{$MAPLOOP}, match={Applied}, replace={Appl.}]
			\step[fieldsource=\regexp{$MAPLOOP}, match={Administrative}, replace={Administ.}]
			\step[fieldsource=\regexp{$MAPLOOP}, match={Approximate}, replace={Approx.}]
			\step[fieldsource=\regexp{$MAPLOOP}, match={Advanced}, replace={Adv.}]
			\step[fieldsource=\regexp{$MAPLOOP}, match={Advances}, replace={Adv.}]
			\step[fieldsource=\regexp{$MAPLOOP}, match={Archives}, replace={Arch.}]
			\step[fieldsource=\regexp{$MAPLOOP}, match={Archive}, replace={Arch.}]
			\step[fieldsource=\regexp{$MAPLOOP}, match={Aeronautics}, replace={Aeronaut.}]
			\step[fieldsource=\regexp{$MAPLOOP}, match={Artificial}, replace={Artif.}]
			\step[fieldsource=\regexp{$MAPLOOP}, match={Aerospace}, replace={Aerosp.}]
			\step[fieldsource=\regexp{$MAPLOOP}, match={Assembly}, replace={Assem.}]
			\step[fieldsource=\regexp{$MAPLOOP}, match={Affective}, replace={Affect.}]
			\step[fieldsource=\regexp{$MAPLOOP}, match={Association}, replace={Assoc.}]
			\step[fieldsource=\regexp{$MAPLOOP}, match={Africa}, replace={Afr.}]
			\step[fieldsource=\regexp{$MAPLOOP}, match={African}, replace={Afr.}]
			\step[fieldsource=\regexp{$MAPLOOP}, match={Astronomy}, replace={Astron.}]
			\step[fieldsource=\regexp{$MAPLOOP}, match={Aircraft}, replace={Aircr.}]
			\step[fieldsource=\regexp{$MAPLOOP}, match={Astronautics}, replace={Astronaut.}]
			\step[fieldsource=\regexp{$MAPLOOP}, match={Algebraic}, replace={Algebr.}]
			\step[fieldsource=\regexp{$MAPLOOP}, match={Astrophysics}, replace={Astrophys.}]
			\step[fieldsource=\regexp{$MAPLOOP}, match={American}, replace={Amer.}]
			\step[fieldsource=\regexp{$MAPLOOP}, match={Atmosphere}, replace={Atmos.}]
			\step[fieldsource=\regexp{$MAPLOOP}, match={Atomic}, replace={At.}]
			\step[fieldsource=\regexp{$MAPLOOP}, match={Atoms}, replace={At.}]
			\step[fieldsource=\regexp{$MAPLOOP}, match={Broadcasting}, replace={Broadcast.}]
			\step[fieldsource=\regexp{$MAPLOOP}, match={Australasian}, replace={Australas.}]
			\step[fieldsource=\regexp{$MAPLOOP}, match={Bulletin}, replace={Bull.}]
			\step[fieldsource=\regexp{$MAPLOOP}, match={Australia}, replace={Aust.}]
			\step[fieldsource=\regexp{$MAPLOOP}, match={Bureau}, replace={Bur.}]
			\step[fieldsource=\regexp{$MAPLOOP}, match={{Automatic }}, replace={Autom.}]
			\step[fieldsource=\regexp{$MAPLOOP}, match={Business}, replace={Bus.}]
			\step[fieldsource=\regexp{$MAPLOOP}, match={Automation}, replace={Automat.}]
			\step[fieldsource=\regexp{$MAPLOOP}, match={Canadian}, replace={Can.}]
			\step[fieldsource=\regexp{$MAPLOOP}, match={Automotive}, replace={Automot.}]
			\step[fieldsource=\regexp{$MAPLOOP}, match={Automobiles}, replace={Automob.}]
			\step[fieldsource=\regexp{$MAPLOOP}, match={Automobile}, replace={Automob.}]
			\step[fieldsource=\regexp{$MAPLOOP}, match={Ceramic}, replace={Ceram.}]
			\step[fieldsource=\regexp{$MAPLOOP}, match={Autonomous}, replace={Auton.}]
			\step[fieldsource=\regexp{$MAPLOOP}, match={Chemical}, replace={Chem.}]
			\step[fieldsource=\regexp{$MAPLOOP}, match={Behavioral}, replace={Behav.}]
			\step[fieldsource=\regexp{$MAPLOOP}, match={Behavior}, replace={Behav.}]
			\step[fieldsource=\regexp{$MAPLOOP}, match={Chinese}, replace={Chin.}]
			\step[fieldsource=\regexp{$MAPLOOP}, match={Belgian}, replace={Belg.}]
			\step[fieldsource=\regexp{$MAPLOOP}, match={Climatology}, replace={Climatol.}]
			\step[fieldsource=\regexp{$MAPLOOP}, match={Biochemical}, replace={Biochem.}]
			\step[fieldsource=\regexp{$MAPLOOP}, match={Clinical}, replace={Clin.}]
			\step[fieldsource=\regexp{$MAPLOOP}, match={Bioinformatics}, replace={Bioinf.}]
			\step[fieldsource=\regexp{$MAPLOOP}, match={Cognitive}, replace={Cogn.}]
			\step[fieldsource=\regexp{$MAPLOOP}, match={Biology, Biological}, replace={Biol.}]
			\step[fieldsource=\regexp{$MAPLOOP}, match={Colloquium}, replace={Colloq.}]
			\step[fieldsource=\regexp{$MAPLOOP}, match={Kolloquium}, replace={Kolloq.}]
			\step[fieldsource=\regexp{$MAPLOOP}, match={Biomedical}, replace={Biomed.}]
			\step[fieldsource=\regexp{$MAPLOOP}, match={Communications}, replace={Commun.}]
			\step[fieldsource=\regexp{$MAPLOOP}, match={Communication}, replace={Commun.}]
			\step[fieldsource=\regexp{$MAPLOOP}, match={Biophysics}, replace={Biophys.}]
			\step[fieldsource=\regexp{$MAPLOOP}, match={Compatibility}, replace={Compat.}]
			\step[fieldsource=\regexp{$MAPLOOP}, match={British}, replace={Brit.}]
			\step[fieldsource=\regexp{$MAPLOOP}, match={Components}, replace={Compon.}]
			\step[fieldsource=\regexp{$MAPLOOP}, match={Component}, replace={Compon.}]
			\step[fieldsource=\regexp{$MAPLOOP}, match={Computational}, replace={Comput.}]
			\step[fieldsource=\regexp{$MAPLOOP}, match={Delivery}, replace={Del.}]
			\step[fieldsource=\regexp{$MAPLOOP}, match={Computers}, replace={Comput.}]
			\step[fieldsource=\regexp{$MAPLOOP}, match={Computer}, replace={Comput.}]
			\step[fieldsource=\regexp{$MAPLOOP}, match={Department}, replace={Dept.}]
			\step[fieldsource=\regexp{$MAPLOOP}, match={Computing}, replace={Comput.}]
			\step[fieldsource=\regexp{$MAPLOOP}, match={Design}, replace={Des.}]
			\step[fieldsource=\regexp{$MAPLOOP}, match={Condensed}, replace={Condens.}]
			\step[fieldsource=\regexp{$MAPLOOP}, match={Detector}, replace={Detect.}]
			\step[fieldsource=\regexp{$MAPLOOP}, match={Conferences}, replace={Conf.}]
			\step[fieldsource=\regexp{$MAPLOOP}, match={Conference}, replace={Conf.}]
			\step[fieldsource=\regexp{$MAPLOOP}, match={Development}, replace={Develop.}]
			\step[fieldsource=\regexp{$MAPLOOP}, match={Congress}, replace={Congr.}]
			\step[fieldsource=\regexp{$MAPLOOP}, match={Differential}, replace={Differ.}]
			\step[fieldsource=\regexp{$MAPLOOP}, match={Consumer}, replace={Consum.}]
			\step[fieldsource=\regexp{$MAPLOOP}, match={Digest}, replace={Dig.}] 
			\step[fieldsource=\regexp{$MAPLOOP}, match={{ of the }}, replace={{ }}]
			\step[fieldsource=\regexp{$MAPLOOP}, match={{ of }}, replace={{ }}]
			\step[fieldsource=\regexp{$MAPLOOP}, match={{Of }}, replace={{ }}]
			\step[fieldsource=\regexp{$MAPLOOP}, match={{ on }}, replace={{ }}]
			\step[fieldsource=\regexp{$MAPLOOP}, match={{ On }}, replace={{ }}]
			\step[fieldsource=\regexp{$MAPLOOP}, match={{On }}, replace={{ }}]
			\step[fieldsource=\regexp{$MAPLOOP}, match={{ in }}, replace={{ }}]
			\step[fieldsource=\regexp{$MAPLOOP}, match=\regexp{\\\x{26}}, replace={{and}}] 
			\step[fieldsource=\regexp{$MAPLOOP}, match={{, and }}, replace={{, }}]
			\step[fieldsource=\regexp{$MAPLOOP}, match={{, And }}, replace={{, }}]
			\step[fieldsource=\regexp{$MAPLOOP}, match={{ and }}, replace={{ }}]
			\step[fieldsource=\regexp{$MAPLOOP}, match={{ And }}, replace={{ }}]
			\step[fieldsource=\regexp{$MAPLOOP}, match={{ In }}, replace={{ }}]
			\step[fieldsource=\regexp{$MAPLOOP}, match={{In }}, replace={{ }}]
			\step[fieldsource=\regexp{$MAPLOOP}, match={{ the }}, replace={{ }}] 
			\step[fieldsource=\regexp{$MAPLOOP}, match={{ The }}, replace={{ }}] 
			\step[fieldsource=\regexp{$MAPLOOP}, match={{The }}, replace={}] 
			\step[fieldsource=\regexp{$MAPLOOP}, match={{ for }}, replace={{ }}] 
			\step[fieldsource=\regexp{$MAPLOOP}, match={First}, replace={1st}]
			\step[fieldsource=\regexp{$MAPLOOP}, match={Second}, replace={2nd}]
			\step[fieldsource=\regexp{$MAPLOOP}, match={Third}, replace={3rd}]
			\step[fieldsource=\regexp{$MAPLOOP}, match={Fourth}, replace={4th}]
			\step[fieldsource=\regexp{$MAPLOOP}, match={Fifth}, replace={5th}]
			\step[fieldsource=\regexp{$MAPLOOP}, match={Sixth}, replace={6th}]
			\step[fieldsource=\regexp{$MAPLOOP}, match={Seventh}, replace={7th}]
			\step[fieldsource=\regexp{$MAPLOOP}, match={Eighth}, replace={8th}]
			\step[fieldsource=\regexp{$MAPLOOP}, match={Ninth}, replace={9th}]
			\step[fieldsource=\regexp{$MAPLOOP}, match={Tenth}, replace={10th}]
			\step[fieldsource=\regexp{$MAPLOOP}, match={Eleventh}, replace={11th}]
			\step[fieldsource=\regexp{$MAPLOOP}, match={Twelfth}, replace={12th}]
			\step[fieldsource=\regexp{$MAPLOOP}, match={Thirteenth}, replace={13th}]
			\step[fieldsource=\regexp{$MAPLOOP}, match={Fourteenth}, replace={14th}]
			\step[fieldsource=\regexp{$MAPLOOP}, match={Fifteenth}, replace={15th}]
			\step[fieldsource=\regexp{$MAPLOOP}, match={Sixteenth}, replace={16th}]
			\step[fieldsource=\regexp{$MAPLOOP}, match={Seventeenth}, replace={17th}]
			\step[fieldsource=\regexp{$MAPLOOP}, match={Eighteenth}, replace={18th}]
			\step[fieldsource=\regexp{$MAPLOOP}, match={Nineteenth}, replace={19th}]
			\step[fieldsource=\regexp{$MAPLOOP}, match={Twentieth}, replace={20th}]
		}
	}
}
\usepackage{xpatch}
\usepackage{xstring}

\DeclareSourcemap{
	\maps[datatype=bibtex]{
		\map{
			\step[fieldsource=langid, match=\regexp{\A(n)?((swiss)?german|austrian)\Z}, final]
			\step[fieldset=language, fieldvalue={(in German)}]
		}
		\map{
			\step[fieldsource=langid, match=pinyin, final]
			\step[fieldset=language, fieldvalue={(in Chinese)}]
		}
		\map{
			\step[fieldsource=langid, match=japanese, final]
			\step[fieldset=language, fieldvalue={(in Japanese)}]
		}
		\map{
			\step[fieldsource=langid, match=ko, final]
			\step[fieldset=language, fieldvalue={(in Korean)}]
		}
	}
}

\xpatchbibdriver{inproceedings}					 	
	{\usebibmacro{publisher+location+date}}			
	{\IfSubStr{\strfield{booktitle}}{\strfield{year}}{\usebibmacro{publisher+location}}{\usebibmacro{publisher+location+date}}} 
	{} 
	{\typeout{There was an error patching biblatex-ieee (specifically, ieee.bbx's @inproceedings driver)}} 

\newbibmacro*{publisher+location}{%
	\printlist{location}%
	\iflistundef{publisher}
	{\setunit*{\addcomma\space}}
	{\setunit*{\addcolon\space}}%
	\printlist{publisher}%
	\newunit}

\xpatchbibdriver{online}
{\printtext[parens]{\usebibmacro{date}}}
	{\iffieldundef{year}
		{}
		{\printtext[parens]{\usebibmacro{date}}}}
	{}
	{\typeout{There was an error patching biblatex-ieee (specifically, ieee.bbx's @online driver)}}

\DeclareSourcemap{
	\maps[datatype=biber]{
		\map{
			\step[fieldsource=note, final]
			\step[fieldset=addendum, origfieldval, final]
			\step[fieldset=note, null]
		}
	}
}

\DeclareBibliographyDriver{standard}{%
	\usebibmacro{bibindex}%
	\usebibmacro{begentry}%
	\usebibmacro{maintitle+title}%
	\setunit{\addcomma\addspace}%
	\usebibmacro{publisher+type+number}%
	\setunit{\labelnamepunct}\newblock
	\IfSubStr{\strfield{number}}{\strfield{year}}{}{\usebibmacro{date}} 
	\newunit\newblock
	\usebibmacro{finentry}%
}

\newbibmacro*{publisher+type+number}{%
	\printtext{%
		\printlist{publisher}
		\iflistundef{publisher}
		{\space}{}%
		\iffieldundef{type}
		{Standard}
		{\printfield{type}}
		\printfield{number}
	}%
}

\xpatchbibdriver{report}					 	
	{\usebibmacro{date}}			
	{\IfSubStr{\strfield{number}}{\strfield{year}}{}{\usebibmacro{date}}} 
	{} 
	{\typeout{There was an error patching biblatex-ieee (specifically, ieee.bbx's @report driver)}} 

\DeclareBibliographyDriver{software}{%
	\usebibmacro{bibindex}%
	\usebibmacro{begentry}%
	\usebibmacro{maintitle+title}%
	\setunit{\addcomma\addspace}%
	\usebibmacro{version+year}%
	\setunit{\labelnamepunct}\newblock
	\usebibmacro{author}%
	\setunit{\addcomma\addspace}%
	\newunit\newblock
	\usebibmacro{url+urldate}%
	\newunit\newblock
	\usebibmacro{finentry}%
}

\newbibmacro*{version+year}{%
	\printtext{%
		\iffieldundef{version}{\iffieldundef{year}{}{\printfield{year}}}{\printfield{version}}
		\printfield{number}
		\printlist{publisher}
		\iflistundef{publisher}{\space}{}%
	}%
}

\usepackage{cleveref}
\crefname{figure}{Fig.}{Fig.}
\crefname{equation}{}{}
\Crefname{equation}{Equation}{Equations}

\usepackage{siunitx}
\DeclareSIUnit\foot{ft}

\makeatletter
\g@addto@macro\@floatboxreset{\def\@tempa{table}\ifx\@captype\@tempa\footnotesize\fi}
\makeatother

\def\BibTeX{{\rm B\kern-.05em{\sc i\kern-.025em b}\kern-.08em
    T\kern-.1667em\lower.7ex\hbox{E}\kern-.125emX}}

\newcommand{\orcid}[1]{\href{https://orcid.org/#1}{\includegraphics[width=8pt]{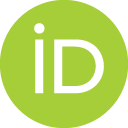}}}

\usepackage{hologo}
\usepackage{listings}
\newif\ifanonymous
 \anonymousfalse

\newif\iftitlepage
\titlepagefalse

\newif\ifabstractonly
\abstractonlyfalse

\ifabstractonly\titlepagefalse\fi
\iftitlepage\anonymousfalse\fi
\ifanonymous\doubleblindtrue\fi

\newif\ifmaintext
\maintexttrue
\iftitlepage\maintextfalse\fi
\ifabstractonly\maintextfalse\fi

\begin{document}

\onecolumn
\begin{center}{\LARGE\bfseries Preprint Notice\par}\end{center}
\vspace{2ex}
\noindent This work has been submitted for possible publication. Copyright may be transferred without notice, after which this version may no longer be accessible.\par
\vspace{2ex}
\thispagestyle{empty}
\setcounter{page}{0}
\clearpage
\twocolumn

\clearpage

\begin{frontmatter}

\title{
Safety Blind Spot in Remote Driving: Considerations for Risk Assessment of Connection Loss Fallback Strategies
}

\ifanonymous\else
\author{Leon Johann Brettin\corref{cor1}\,\orcid{0009-0007-2796-0860}}
\ead{l.brettin@tu-braunschweig.de}
\author{Niklas Braun\,\orcid{0000-0002-9312-8446}}
\author{Robert Graubohm\,\orcid{0000-0002-6682-4788}}
\author{Markus Maurer\,\orcid{0000-0002-5357-9701}}

\affiliation{organization={Institute of Control Engineering, Technische Universit\"at Braunschweig},
            postcode={38106},
            city={Braunschweig},
            country={Germany}}

\cortext[cor1]{Corresponding author.}
\fi

\iftitlepage\else   
\begin{abstract}

As part of the overall goal of driverless road vehicles, remote driving is a major emerging field of research of its own.
Current remote driving concepts for public road traffic often establish a fallback strategy of immediate braking to a standstill in the event of a connection loss.
This may seem like the most logical option when human control of the vehicle is lost.
However, our simulation results from a total of $\sim 186000$ start scenes based on naturalistic traffic scenarios indicate high collision rates for any immediate substantial deceleration to a standstill in urban settings.
We show that such a fallback strategy can result in a SOTIF-relevant hazard, raising the question whether this design decision is acceptable.
Therefore, from a safety perspective, we call this problem a \emph{safety blind spot}, as safety analyses in this regard seem to be very rare. 

In this article, we first present a simulation on a naturalistic dataset that indicates that collisions are plausible in a substantial share of the simulated start scenes.
Second, we discuss the severity of the resulting potential rear-end collisions and provide an even more severe example by including a large commercial vehicle in the potential collision.

\end{abstract}

\ifabstractonly\else
\fi
\fi   

\end{frontmatter}

\ifmaintext

\section{Introduction}

The research on teleoperated vehicles and as a subclass \mbox{remote} driving has seen increased research interest in recent years:
There are legal developments (see, e.g. German Traffic Law~\cite[§~1d]{StVG_1d_BGB}, German Regulation \cite{StVFernLV2025} or new UK BSI-Flex~\cite{BSI_1886}), start-up \mbox{companies} (see, e.g. Vay\footnote{\url{https://vay.io}}, Elmo\footnote{\url{https://www.elmoremote.com}}, Fernride\footnote{\url{https://www.fernride.com}}, Voysis\footnote{\url{https://www.voysys.se}}), market opportunities (see, e.g. \cite{Kelkar2025_McKinsey}) and new research questions (see, e.g. \cite{WG_Teleoperation2024_en}) in this area, motivating new system designs and publications.

We observe that many current teleoperation initiatives and regulatory efforts \emph{assume} that emergency stopping in the event of major system failures is a design decision that, at least in urban operating environments, generally does not pose unreasonable risk. 
In this context, connection loss would be one of the failure events that would cause remote-driven vehicles to come to an immediate stop. 
In this article we take a closer look at the probability and potential consequences of collisions with closely following vehicles in urban traffic scenarios and find that they can be significant. 

Depending on the specific operating environment (e.g., geographic area, network load, time of day), the stability of the connection can be improved, but a loss cannot be completely excluded. In the absence of evidence that a connection loss is impossible, a safety concept has to assume that it can occur. 
Since the consequences of this failure are rarely addressed in published safety analyses and, if addressed, not further analyzed, we consider this to be a safety blind spot.

The contribution of this paper is to identify the handling of connection loss in remote driving as an under-examined safety blind spot.
We argue that immediate emergency braking as a connection-loss fallback constitutes a safety hazard whose acceptability is questionable, supported by a naturalistic-data plausibility study and a hazard analysis based on ISO 21448.

We structure the argumentation of this article around two main arguments based on the definition of risk in ISO~26262, which is defined as ``combination of the probability of occurrence of harm [...] and the severity [...] of that harm [...]''~\cite{ISO26262}.
\textit{First} we look at the \emph{probability of occurrence} by conducting a \mbox{simulation} using naturalistic driving data:
We simulate \mbox{braking} as a fallback strategy for the rear vehicle with two driver models, one based on the \mbox{\emph{intelligent driver model}}~\cite{Treiber2000} and the other based on a simplified braking model that applies full braking after a fixed reaction time.
The simulation is configured purely longitudinal with start scenes and initial parameters based on the naturalistic real-world \emph{uniD} dataset~\cite{Krajewski2018, Bock2019, Krajewski2020, Moers2022}.
It should be noted that the simulation is not intended to estimate real-world collision frequencies. 
It only serves as a plausibility argument showing that under naturalistic following distances and reasonable assumptions about human reactions, rear-end collisions after connection loss are possible.
This is sufficient for the probability argument, showing that this hazard is relevant for risk assessment.
Quantifying the frequency with further simulations is beyond the scope of this article.

\textit{Second}, we look at the potential \emph{severity} by assessing potential hazards associated with the analyzed fallback strategy.
For that, we will discuss the same scenario based on ISO~21448~\cite{ISO21448} considerations, as this type of collision could result in a potentially high severity.

It is important to note that the purpose of this paper is to provide a practical example of why an emergency deceleration in the event of a connection loss can be considered unsafe. 
We acknowledge that the connection component is widely considered critical in a remote vehicle. 
However, when it comes to discussing how to handle connection loss, this article presents an argument about the risks involved in stopping in a lane in such circumstances.

This article is intended for developers, safety engineers, system designers, regulators, and other stakeholders planning to introduce remotely operated vehicles into regular traffic:
When considering this safety blind spot from a \mbox{system} design perspective, identifying a lack of an \mbox{adequate} fallback strategy can potentially avoid unknown costs, \mbox{injuries}, or further damage.
Therefore, it should be beneficial to design a sound fallback strategy as early as the concept phase of development, rather than hoping that the connection will break very rarely and never under the wrong circumstances.
However, the purpose of this article is not to resolve this blind spot, but to highlight it and to provide a reference point for an argument against such braking strategies, in the hope of stimulating discussion about a possible specification for a solution.

One of the main motivations for this work was the \mbox{collaborative} work done in the Working Group ``Research Needs in Teleoperation'' \cite{WG_Teleoperation2024_en}\ifanonymous\else, in which some of the authors of this article participated\fi.
In this report, research questions on different research areas of teleoperation have been collected. 
One of the questions raised in the first cluster, ``Vehicle, area of operation and functional safety'', was what would happen in the event of a loss of connectivity in a teleoperated vehicle, which motivated the work in this article.

The remainder of the article is therefore structured as follow: In \Cref{sec: terminology}, we present terminology and definitions in the context of remote driving.
Related work is presented and discussed in \Cref{sec: related work}.
A discussion of potential fallback options is provided in \Cref{sec: discussion strategies}.
The analysis and simulation is described in \Cref{sec: analysis}, followed by the risk assessment in \Cref{sec: risk assessment}.
The results of the simulation and assessment are then discussed in \Cref{sec: discussion}.
The conclusion can be found in \Cref{sec: conclusion}.


\section{Terminology and Definitions}
\label{sec: terminology}

This Section provides definitions of important terms used in this article.

\subsection*{Teleoperation}

Teleoperation systems according to \textcite{Winfield2009} require three main parts to function: the robot, the communication link, and the operator interface where an operator can \mbox{remotely} control the vehicle.
The robot, in our case a vehicle, must at least have sensors that allow a human to perceive the environment, an interface to access the vehicle's actuators, and a communication device to establish and use a connection to the operator interface.
Central to this work is the communication link, which transports signals between the operator interface and the vehicle. 
When we refer to the connection in this article, we specifically mean this radio link between the operator interface and the vehicle.

\subsection*{Remote Driving}

This article and the following discussion are focused on \emph{remote driving} as defined in SAE~J3016 as ``Real-time performance of part or all of the [Dynamic Driving Task (DDT)] and/or DDT fallback (including, real-time braking, steering, acceleration, and transmission shifting), by a remote driver.''~\cite[Clause~3.24]{J3016}. 
This definition does not require the need for automated driving capability in such a vehicle system, a point also mentioned in \cite{WG_Teleoperation2024_en}. 
This is central to our argument: if no on-board automation can perform the DDT, a connection loss leaves the vehicle without any system able to safely continue the dynamic driving task. 
Immediate braking remains a simple, but as we argue, unsafe, fallback.

We understand teleoperation as a generic term for the research area discussed in this article, with remote driving being a mode of teleoperation.
However, teleoperation can also take other modes, such as \emph{Remote Assistance}~\cite[Clause~3.23]{J3016} or even more specific modes further discussed by \textcite{Majstorovic2022}, which is important to be aware of, but not part of this article.

\subsection*{Controllability}
For this article we follow the definition of \emph{Controllability} given in ISO~26262 \cite[Clause~3.25]{ISO26262} as the ``ability to avoid a specified harm [...] or damage through the timely reactions of the persons involved, possibly with support from external measures [...]''.
The same term is used in ISO~21448~\cite{ISO21448} as one of the classification indicator for harm. 

A definition is important here, because we argue that controllability is absent in the case of a connection loss.
When the connection is lost, neither the remote operator, who is no longer connected, nor the driver of a following vehicle, who cannot anticipate the sudden braking, is able to react in time to avoid the resulting harm.


\section{Related Work}
\label{sec: related work}

This Section reviews prior work on teleoperation with a focus on the communication link, and discusses the relation to our contribution. 
While the criticality of the connection, including latency and connection loss, is widely recognized, we show that the specific hazard we address, rear-end collisions caused by the fallback itself, has received little attention.
The research questions collected in \cite{WG_Teleoperation2024_en}, which was one of the motivations for this work, already point to the open problem of how a teleoperated vehicle should behave when the connection is lost.

 The German government recently released a regulation for remote driving (StVFernL)~\cite{StVFernLV2025} focusing on vehicles that are solely remotely driven. 
This regulation appears to be inspired by the German national regulation for SAE Level 4 vehicles (AfGBV) \cite{Germany_AFGBV_2022_eng}. 
It allows vehicles to be driven remotely within a permitted operational domain, provided that the requirements specified in the document for the owner, the responsible department and the remote operator are met.
Relevant to our argument, the regulation requires a system for minimal risk maneuver that is distinct from an automatic emergency braking function, but it does not specify what this system must be capable of.
We use this article to examine the limitations of immediate braking to a standstill as such a minimal-risk maneuver in the analyzed scenario, showing that the problem is more complex than this open specification suggests.

The UK Law Commission identifies rear-end collisions as a potential problem~\cite[A.20--A.23, A.35]{UKLawCommission2023}.
It recommends that operators of remote-driven vehicles address the risk mitigation system in case of a connection loss in their safety case, and notes that a vehicle stopping in its lane after a connection loss might be struck from behind.
However, it leaves it to the developer to show that the system is safe enough in its context.

The communication link is a critical and externally dependent component of a remote-driven vehicle.
Unlike the other components of the system, which are part of the vehicle itself, the link depends on external factors beyond the system's control, and its quality varies with the operating environment, a dependency that does not exist for an onboard driver.
The literature shows that with higher latency the controllability degrades \cite{Zhang2020} (following \cite{Sheridan1963, Luck2006, Chen2007, Davis2010, Gnatzig2013, Bodell2016, Liu2017})
and propose mitigations and solutions for latency problems, such as 
connectivity-aware routing~\cite{Kay1995}, latency dependent speed adaptation~\cite{Neumeier2019}, or latency optimized hardware~\cite{Georg2020}. 
These approaches reduce the effects of a degraded connection, but they cannot eliminate the most severe failure mode: the complete loss of connection, which we discuss in this article. 

\textcite{Zhang2020} describes what a remote-operated system must provide: awareness of a potential connection loss, the ability to go into a ``safe mode'' as a precaution, and, most importantly for this article, the ability for the vehicle to reach a ``safe mode'' without the operator.
\citeauthor{Tang2014} likewise identify the choice of an appropriate safety concept in case of a connection loss as one of the main challenges in teleoperated vehicles \cite[1399]{Tang2014}.
This last requirement is the problem we want to address in this work: whether immediate braking can serve as such a ``safe mode'', and what follows if it cannot.

Efforts are already being made to address this problem. For example, \cite{Diermeyer2011, Tang2014, Hoffmann2022, Brecht2024_Risk} are introducing the concept of a \emph{free corridor} where an area in front of the vehicle has to be kept free from obstacles by the operator in order to allow the vehicle to decelerate to a standstill within the set corridor.
In that case, maintaining the corridor as a free space is a key task for the operator.
The same group also identified network connection loss with the help of System-Theoretic Process Analysis (STPA) as a causal factor, describing the network as ``one of the most critical parts within the whole teleoperation system''~\cite[453]{Hoffmann2021}.

The free corridor approaches address the space in front of the vehicle and the risk of braking or driving into something ahead.
This article instead focuses on the hazard \emph{behind} the remote-driven vehicle: a closely following vehicle cannot anticipate the sudden braking triggered by a connection loss, and may therefore collide with it.

Emergency braking maneuvers that cannot be anticipated can lead to collisions, even in lower-speed areas such as an urban environment (e.g., \cite{NHTSA2007}, \cite{Distner2009} following \cite{Avery2008}, \cite{Beyerer2019}, \cite{Bendrick2025}). Braking is therefore not inherently a safe response: under certain conditions, triggering emergency braking may even increase the severity of harm \cite{Heck2015, Beyerer2019} (for example, false-triggers of automatic emergency braking are simulated for closed freeways in \cite{Faerber2005}). This is exactly the situation we examine, where the braking of the remote-driven vehicle is unforeseeable to the following driver.

\textcite{Tandon2026} recently published a preprint that likewise highlights the importance of considering minimal risk maneuvers for autonomous vehicles, showing that an improperly stopped autonomous vehicle can lead to unsafe interactions. They show this by analyzing incident reports in which fallback behaviors of autonomous vehicles resulted in obstructed traffic and related issues.

For the simulation, we present in \Cref{sec: analysis}, we use the \emph{uniD} dataset ~\cite{Krajewski2018, Bock2019, Krajewski2020, Moers2022}, which recorded naturalistic trajectories on a straight, urban road.
This setting matches our scenario of a vehicle closely following another with no curves, roundabouts or intersections, supporting a purely longitudinal analysis of a braking fallback.

The behavior of human drivers on the \emph{uniD} dataset is itself an object of study. \textcite{Certad2023} use the same \emph{uniD} dataset with the IEEE~2846-2022 standard \cite{IEEE2846-2022} to derive assumptions about kinematic values for ``reasonably foreseeable behavior'' as starting conditions for testing vehicles equipped with an automated driving system. This is relevant to our work in two ways: it shows that the \emph{uniD} dataset is established for analyzing naturalistic road-user behavior, and it supports grounding assumptions about human reactions, such as those in our simulation, in naturalistic data.

A considerable body of work thus addresses teleoperation and its safety implications.
To the best of our knowledge, however, no prior work exists that uses naturalistic data to assess the collision risk of the braking fallback after connection loss for a specific operational environment, nor examines the resulting rear-end hazard for vehicles behind the remote-driven vehicle.


\section{Potential deceleration strategies on connection loss for Remote Driving}
\label{sec: discussion strategies}

\begin{figure}
    \centering
    \includegraphics[width=0.99\columnwidth]{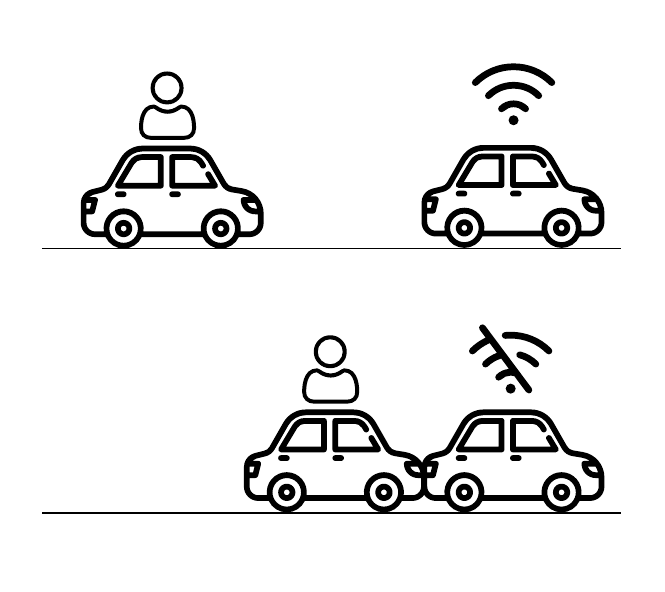}
    \caption{
    A human-driven vehicle is following a remote-driven vehicle. 
    A loss of connection occurs in the remote-driven vehicle, leading to a fallback strategy of braking which could result in a collision.
    }
    \label{fig: overview}
\end{figure}

When a remote-driven vehicle loses its connection, it must react without a driver performing the dynamic driving task. 
This Section discusses potential deceleration strategies available for the remote-driven vehicle in such situations.
Throughout, we consider the scenario of a human-driven vehicle closely following the remote-driven vehicle (see \cref{fig: overview}); since closely following another vehicle is already an everyday situation, this will be equally common once remote-driven vehicles operate more often in public traffic.

Continuing to drive would involve an unreasonably high risk and is therefore unsafe, because neither a driver nor a remote driver is performing the dynamic driving task and the vehicle would continue without any navigation or guidance.
The vehicle system therefore has to apply a fallback strategy and brake.
Regardless of whether the steering wheel is set straight or the steering angle is fixed, lane departure remains imminent.
Braking itself can be realized in different ways: either with maximum deceleration to reach a standstill within the minimal stopping distance, or with a more moderate deceleration.
In the latter case the stopping distance is longer, but a following human-operated vehicle has more time and distance available to avoid a collision, which, as the next Subsection discusses, can be decisive.

\subsection{Emergency Stop}
The first possible reaction is to perform an emergency stop when the connection is lost. At first, this might seem the most natural because it would probably be what a human would try to do.
If a human driver were to lose sight and control from one second to the next, braking at the physical limits would reduce the probability of hitting something or someone in the direction of motion.
As will be discussed in the next Subsection, a naturalistic emergency brake is usually not at the physical limits, but for the theoretical thinking of hitting the brake as hard as possible, it seems like a natural idea.

\subsubsection*{Humans cannot foresee such a stop}
A human operating the closely following vehicle behind the stopping remote-driven vehicle has no chance to anticipate the situation, because a connection loss is something that cannot be seen from the outside, such as a red traffic light or a child or adult pedestrian crossing the street in front of the vehicle ahead. 
Finding an analogy for human driving in a similar setting is not easy: In fog or heavy rain, sudden strong deceleration has a plausible, perceptible cause for the surrounding traffic, so following drivers can expect decelerations and adapt their driving style in anticipation. 
Where it is not perceptible, for instance when a stone shatters the windscreen, the driver of the vehicle still remains able to judge the situation and to decelerate in a controlled manner.
A connection loss provides neither.
But in the scenario of this article, the following human driver has a fundamentally different understanding of the situation than the remote-driven vehicle:
The human driver will not anticipate the connection loss before it occurs and must react to the sudden deceleration, braking lights and possibly hazard warning lights of the remote-driven vehicle ahead.
This unpredictability for the human driver increases the probability of a rear-end collision compared to sudden braking events in today's traffic.

\subsection{Moderate Deceleration}
Another possible reaction is to use moderate deceleration instead of full braking, resulting in a longer stopping distance. 
This increases the probability of colliding with an object in the vehicle's direction of motion compared to full braking. 

For the sake of simplicity, we make no assumptions about the vehicle's \mbox{capability} of lateral guidance. Hence a longer stopping distance may significantly increase the probability of exiting the lane boundaries as well. 

As mentioned in \Cref{sec: related work}, researchers from TU Munich~\cite{Diermeyer2011, Tang2014, Hoffmann2022} introduce the \emph{free corridor}, assuming some remaining lateral guidance capability on the vehicle side. 
Here, a path is projected to the operator that the vehicle would follow in the event of a connection loss. 
The path must be kept free by the operator, so that the vehicle has a collision-free path to follow and to brake within if a connection loss would occur, assuming that indeed no objects are in the path.

One open question is which time horizon the \emph{free corridor} needs to cover:
If it is too long, the operator might not be able to keep it clear of objects or might be perceived to drive unreasonably slowly.
In complicated situations such as certain curvy roads, lane modifications due to construction zones, or parked vehicles occupying parts of the traffic lane, it might not always even be feasible for the operator to keep the corridor free.
If the corridor is too short, the vehicle has to apply almost full braking to stop within the short corridor and still have the problem of rear-end collisions. 

There is also the problem that future movements of other objects through the corridor are not initially mapped.
For example, if people are occluded, the remote operator is unlikely to make a speed adjustment that sufficiently shortens the stopping distance if a person suddenly steps into the lane.
This is also a problem for human drivers and automated driving systems, as they, too, may not detect occluded pedestrians when driving at a speed appropriate for the traffic conditions.
A discussion about this topic can be found in \textcite{Graubohm2023_Occlusion}.

The rear-end collision problem is demonstrated in the next Section using a simulation study.

\subsection*{Why not use ADAS lateral and longitudinal guidance capabilities as a fallback?}
An obvious question is whether existing ADAS could simply serve as the fallback.
They cannot.
Automatic emergency braking is designed only to mitigate harm in imminent collisions, not to perform the dynamic driving task; it can reduce the consequences of accidents but does not guarantee collision avoidance (\cite{Rieken2016}, \cite{Reschka2017} following \cite{Daimler2009, vonHolt2004}).
Moreover, activating automatic emergency braking in the absence of an unavoidable-collision criterion can itself be viewed as hazardous \cite[1198]{Rieken2016}.
Level 1 and Level 2 features are equally unsuitable: by definition they require a driver to monitor the environment and the assistance system \cite{J3016}, and no such supervisor is present once the operator is disconnected.
Neither active safety systems nor Level 1/2 driving automation are therefore appropriate as a fallback for remote driving.


\section{Analysis based on the uniD dataset}
\label{sec: analysis}

To illustrate the aforementioned problems, we performed an analysis on the ``naturalistic'' \emph{uniD} dataset~\cite{Krajewski2018, Bock2019, Krajewski2020, Moers2022}, which contains drone recordings of vehicles and pedestrians on a straight road section in an urban area.
Besides this dataset, there are four other datasets that are recorded in the same way: \emph{highD}, a dataset focusing on a highway section \cite{Krajewski2018}, \emph{inD}, focusing on an intersection \cite{Bock2019}, \emph{rounD}, focusing on a roundabout \cite{Krajewski2020} and \emph{exiD}, focusing on a highway exit \cite{Moers2022}.
All these datasets are recorded top-down with a drone over the selected road section and are labeled similarly. 

The simulated scenario contains the mentioned scenario of a human-driven vehicle following a remote-driven vehicle on a straight road. 
The simulation starts with a connection loss of the remote-driven vehicle, which leads to a fallback strategy of braking.
By choosing different parameters for the reaction time for the human-driven vehicle and braking behavior of the human-driven and remote-driven vehicle, it is checked whether the human-driven vehicle collides with the remote-driven vehicle.

In the overall structure of this article, this can be seen as looking at the \emph{probability of occurrence} of such a rear-end collision.
The simulation can be seen here as a plausibility check to obtain a non-negligible probability of a rear-end collision occurring with the fallback strategy of braking.

This Section is structured as follows: 
\Cref{sec: model prequisites} deals with the models used and their prerequisites.
The parameterization of the models based on the relevant literature is also discussed in this part.
In \Cref{sec: data preprocessing}, the preprocessing steps performed to obtain naturalistic scenes of vehicles following each other from the dataset are discussed.
\mbox{\Cref{sec: simulation}} describes the simulation and \Cref{sec: limitations} discusses its limitations.
The technical results of our analysis are presented in \Cref{sec: results}.

The overall results and the acceptability of the findings will be used in the discussion in \Cref{sec: risk assessment} and \Cref{sec: discussion}.

\subsection{Model Prerequisites}
\label{sec: model prequisites}

\begin{figure}
    \centering
    \includegraphics[width=0.99\columnwidth,page=1]{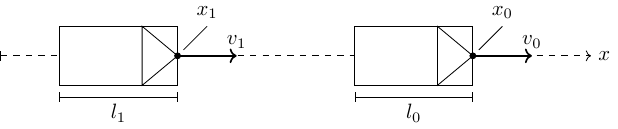}
    \caption{
    One-dimensional simulation of a pair of vehicles on a straight road. The remote-driven vehicle is given index zero and the following human-operated vehicle is given index one. Each vehicle has a position and velocity state and length parameter.
    }
    \label{fig:simulation_sketch}
\end{figure}

We implemented two different longitudinal driver models for our simulation.
We call the first a \emph{sudden braking model} which applies an immediate maximum deceleration \mbox{$\dot{v}^\mathrm{min} \leq 0$} after a reaction time $\tau_r$. The second driver model is the \emph{intelligent driver model} introduced by \citeauthor{Treiber2000}~\cite{Treiber2000}. 
We implemented adjustments from \textcite{Albeaik2022} to deal with deceleration values going to $-\infty$ in the model and unwanted reversing of simulated vehicles.
The main part of the \emph{intelligent driver model} as we use it is shown in \cref{eq: IDM 1,eq: IDM 2,eq: IDM 3}:
\begin{align}
    \beta^\mathrm{free} &= \left(\frac{v}{v_0}\right)^\delta \label{eq: IDM 1}\\
    \beta^{\neg\mathrm{free}} &= \left( \frac{ s_0 + v T + \frac{v d}{2 \sqrt{a b }}}{s} \right) ^2 \label{eq: IDM 2}\\
    \dot{v} &= a \cdot ( 1 -\beta^\mathrm{free} - \beta^{\neg\mathrm{free}})
    \label{eq: IDM 3}
\end{align}
where 
\begin{itemize}
    \item $\beta^\mathrm{free}$ is the term governing the free road behavior;
    \item $\beta^{\neg\mathrm{free}}$ is the term governing the following of a vehicle in front (the lead vehicle);
    \item $\dot{v}$ is the resulting commanded acceleration;
    \item $v$ is the current speed of the considered vehicle;
    \item $v_0$ is the desired speed;
    \item $\delta$ is an arbitrary tuning parameter exponent used by the \emph{intelligent driver model} that provides a degree of freedom to adapt the model's characteristics to observed driver behavior;
    \item $s$ is the actual spacing between considered and lead vehicle;
    \item $s_0$ is the desired minimum spacing between considered and the lead vehicle;
    \item $T$ is the desired time headway to the lead vehicle;
    \item $d$ is the distance between the considered vehicle and lead vehicle;
    \item $a$ is the maximum acceleration commanded by the simulated driver;
    \item $b$ is a tuning parameter related to the desired deceleration.
\end{itemize}
For comparison: The algorithm of \textcite{Treiber2000} has an additional summand $s_1 \sqrt{v/v_0}$ inside the numerator in~\cref{eq: IDM 2}, where \citeauthor{Treiber2000} recommend, for simplicity, to set $s_1$ to zero which is what we did.

As mentioned before, we implemented the modifications from \textcite{Albeaik2022} limiting $v$ and $\dot{v}$ to 
\begin{align}
v_{\mathrm{lim}} &= \max{}(v, 0) \\
\dot{v}_{\mathrm{lim}} &= \max{}(\dot{v}, \dot{v}^\mathrm{min})
\end{align}
where $\dot{v}^\mathrm{min}$ is the maximum deceleration (or minimum \mbox{acceleration}).

\begin{table}[ht]
    \centering
    
    \caption{\textbf{Model parametrization}}
    \begin{tabular}{lcccc}
         \textbf{Description} & \textbf{Variable} & \textbf{Value} & \textbf{Unit} & \textbf{Justification}\\
         \hline
         Maximum acceleration& $a$ &\num{0.73} & \unit{\metre\per\second\squared} & \textbf{*1}\\
         Desired deceleration& $b$ &\num{-1.67}& \unit{\metre\per\second\squared} & \textbf{*1}\\
         Maximum deceleration& $\dot{v}^\mathrm{min}$ & \num{-3.41} & \unit{\metre\per\second\squared} & \textbf{*2}\\
         Desired speed& $v_0$  & \num{50} & \unit{\kilo\metre\per\hour}& \textbf{*3}\\
         Desired time headway& $T$ &  \num{1.6}& \unit{\second}& \textbf{*1}\\
         Minimum spacing& $s_0$ &  \num{2} & \unit{\metre}& \textbf{*1}\\
        Length of the vehicle& $d$ & \num{5} & \unit{\metre}& \textbf{*1}\\
        Acceleration exponent& $\delta$ & \num{4}& & \textbf{*1}\\
        &\\
    \end{tabular}
    \begin{tabular}{c|l}
        & The aforementioned values are required as model parameters \\ & for the selected model. The variable names were derived \\ & from \textcite{Treiber2000}.\\
        \textbf{*1} & \citeauthor{Albeaik2022}~\cite{Albeaik2022} state that these are ``typical and meaningful'' \\ &values according to \citeauthor{Treiber2000}~\cite{Treiber2000}. \\
        \textbf{*2} & \citeauthor{Wood2021}~\cite{Wood2021} state that current standards use a \\& deceleration value of $\qty{11.2}{\foot\per\second\squared}  \approxeq  \qty{3.41}{\metre\per\second\squared}$. \\&
        Since this is debatable, a discussion can be found in the text. \\
        \textbf{*3} & Maximum speed limit on the selected street.
    \end{tabular}
    \label{tab: selected values}
\end{table}

Since one of the most important variables for analysing braking behavior for a human driver is the reaction time, we added a ring buffer to the model, which sets a simplified fixed-time delay of the commanded acceleration signal.

\subsubsection*{Discussion about chosen deceleration}
The values listed in \Cref{tab: selected values} are used for calculation within the \emph{intelligent driver model}.
One could argue that the desired deceleration is still quite low compared to some of the other deceleration values used in the literature. Depending on which source is used, deceleration values below or above the selected value can be found. 
For example, \textcite{Graubohm2023_Occlusion} (derived from \cite{Vangi2007}) use an emergency deceleration of $\qty{6.5}{\meter\per\second\squared}$ and \textcite{Reschka2017} (derived from \cite{ISO22839}) suggests a maximum deceleration of $\qty{6.0}{\meter\per\second\squared}$.
On the other hand, when looking at common braking behavior, e.g., \textcite{Stolte2019} derive $\qty{3.0}{\meter\per\second\squared}$ for a naturalistic drive on an inner city ring road; \textcite{Certad2023} derived $\qty{2.0145}{\meter \per \second \squared}$ for a ``reasonable foreseeable'' deceleration value; and for a non emergency situation \textcite{Deligianni2017} come to average deceleration values of $\qty{2.42}{\meter\per\second\squared}$ (with outliers reaching up to $\qty{5.17}{\meter\per\second\squared}$).
The difference here is that we are looking for a value that real drivers would use in a normal braking situation.
As stated e.g. by \citeauthor{Rieken2016}~\mbox{\cite[p. 1178-1179]{Rieken2016}}, it cannot be assumed that every driver will brake with the maximum deceleration possible (or brake at all) when emergency braking would be necessary. 
Since this simulation aims to simulate human behavior, using full braking power does not seem to achieve this goal. 
While Forward Vehicle Collision Mitigation Systems (FVCMS) according to ISO 22839~\cite{ISO22839} can mitigate many of the simulated collisions with an average deceleration of up to $\qty{6}{\meter\per\second\squared}$ if there is sufficient clearance and a working detection, it cannot be generally assumed that the human driven vehicle in this scenario can use such functions, as the market share and activation status of such functions is unclear.

Further determination of realistic human deceleration is beyond the scope of this work.
In our opinion, $\qty{3.41}{\metre\per\second\squared}$ given by \textcite{Wood2021} seems to be acceptable as a compromise between emergency braking and a normal situation.
For comparison, we will run a simulation with the given emergency deceleration of $\qty{6.5}{\meter\per\second\squared}$, which we will address later in \Cref{sec: results}.

\hfill

We define the \emph{sudden braking model} to have only one parameter other than the reaction time, which is the \mbox{deceleration} after the reaction time. We set the deceleration to the same value as used for the \emph{intelligent driver model} by setting $\dot{v}^\mathrm{min}=\qty{-3.41}{\meter\per\second\squared}$. For a moderate braking configuration of the \emph{sudden braking model}, we set the deceleration to half of that value. 

We decided to use the \emph{intelligent driver model} to simulate human behavior regarding longitudinal vehicle \mbox{control}. While we acknowledge that the parameterization and validation of this and other driver models pose research questions of their own we consider this driver model and its parameters sufficiently valid to investigate naturalistic start scenes with respect to potential rear-end collisions.

An example simulation run can be seen in \cref{fig:simulation_run}. 
The figure shows two simulations with the same starting parameters, but with different driver models. 
The orange line represents the leading vehicle, which performs a naturalistic emergency stop at the beginning of the simulation, resulting in a complete stop of the vehicle.

We have taken the start scene from the later described preprocessing of the uniD dataset: A lead vehicle drives with a speed of $v_0$ and another vehicle follows the vehicle at a distance of $x_0 - x_1$ and a speed of $v_1$ on a straight in-city route.
From there, an emergency stop of the front vehicle is simulated.

The blue dashed line and the green dotted line represent the following vehicle simulated once with the \emph{intelligent driver model} (IDM) and once with the \emph{sudden braking model} (SBM).
The orange line represents the front \mbox{vehicle}, which is simulated using a \emph{sudden braking model} that starts braking at the beginning of the simulation with $a_0 = \qty{-3.41}{\meter \per \second ^2}$.

As it can be seen in the figure, both rear vehicles start braking after a reaction time of $\qty{0.5}{\second}$, as this was set for this example simulation. 
The \emph{intelligent driver model} does not use maximum deceleration and decelerates over a longer time period behind the already stopped vehicle in front.
The IDM-controlled vehicle then stops at the set minimum spacing behind the front vehicle.

\begin{figure}
    \centering
    \includegraphics[width=0.99\columnwidth,page=2]{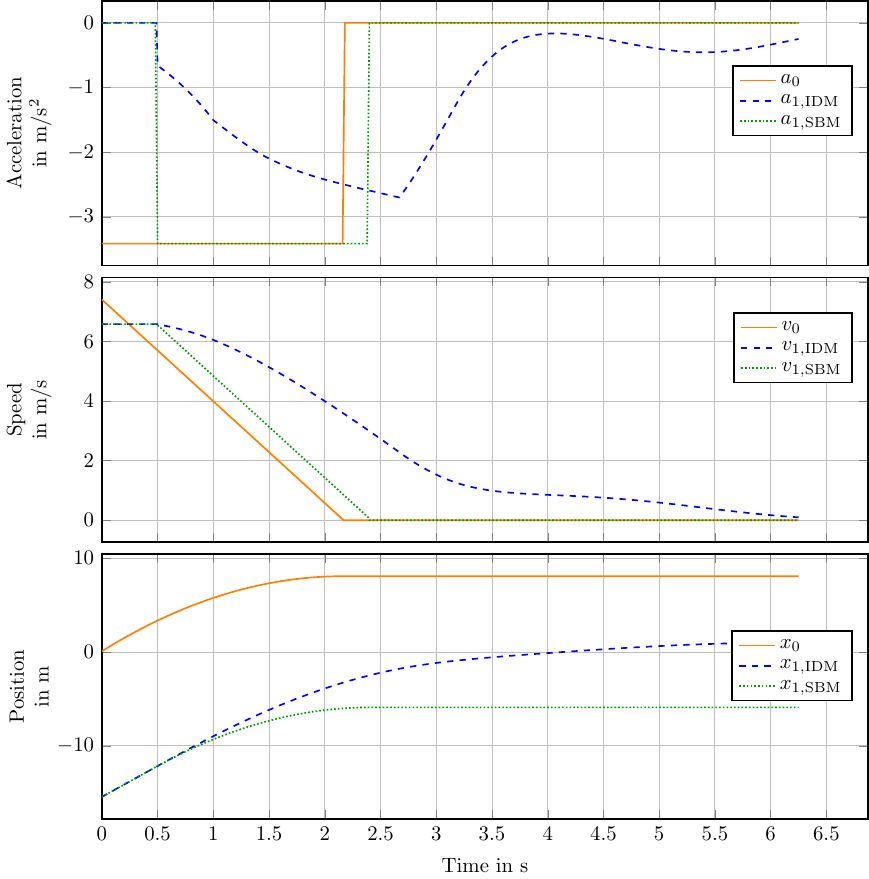}
    \caption{
    Example of a simulation run. The index zero represents the leading vehicle, while the index one denotes the following vehicle. IDM stands for the Intelligent Driver Model, while the SBM represents the Sudden Braking Model. The reaction time for this example is 0.5 s. No collision occurred in this run.
    }
    \label{fig:simulation_run}
\end{figure}

\subsection{Data Preprocessing}
\label{sec: data preprocessing}

As we use naturalistic driving scenes in order to initialize the simulation runs, we preprocess the trajectories from the \emph{uniD} dataset in a pipeline. The pipeline identifies vehicles that follow each other in the recorded stretch of road and yields their respective dynamic states. The filtering aims to prevent false positives and rather strictly discards pairs of vehicles that are not clearly following each other or that may be influenced by other traffic participants. In the following, we explain the filtering steps and parameters used for this study.
It should be noted that the detection of driving maneuvers is a research topic in its own right, and that this processing pipeline is only a simplified version for our purposes where the values and filtering steps had to be chosen based on subjective considerations.

For each frame in the recording the pipeline identifies all pairs of dynamic objects in the scene.
At the start, a frame consists of a list of vehicles with their current information regarding position, speed, type, and others.
We then filter these data objects as follows:

\begin{enumerate}
    \item Make pairs of all vehicle data objects that are in a frame.
    \item Only keep pairs of vehicles that point in roughly the same direction by filtering for a maximum absolute heading difference of $\ang{15}$.
    \item The lead vehicle must be in front of the following vehicle, which we partly assure by limiting the bearing to the lead vehicle from the following vehicle to $\ang{15}$.
    \item Discard pairs where the absolute lateral offset of the lead vehicle from the following vehicle's longitudinal axis is more than $\qty{1}{\metre}$. This accounts for possible maneuver ambiguity, e.g. in case of passing or parking vehicles.
    \item Filter out pairs where other road users are nearby such as \emph{pedestrians, bicycles} or \emph{motorcycles} directly alongside or between the pairing.
    \item Per vehicle pair keep only the pair with the closest lead vehicle.
    \item Discard pair where the following vehicle is following the lead vehicle for less than $\qty{1}{\second}$.
    \item Discard pairs where the following vehicle type is neither \emph{car} or \emph{van} or where the lead vehicle is neither of \emph{a car, a van, a truck, a bus} or \emph{a trailer}. 
\end{enumerate}

\subsection{Simulation}
\label{sec: simulation}

We generate our scenarios as follows:
For each frame of the preprocessed pairs, we define a start scene for the simulation including the initial dynamic states of both the lead and the following vehicles. 
The connection loss of the remote-driven lead vehicle is triggered at the start of the simulation, i.e. in the start scene. The simulation is executed until any termination criterion is reached.
Termination conditions are:  
\begin{enumerate}
    \item reaching maximum simulation duration, 
    \item a collision occurs,
    \item all vehicles have come to a standstill,
\end{enumerate}

The initial scene serves as the starting point for subsequent scenes, which are developed with the help of the simulation. 
With that, the sequence of scenes can be seen as a scenario as defined in \textcite{Ulbrich2015}.

As shown in \Cref{tab: collision rate 1,tab: collision rate 2}, the reaction time $\tau_r$ and the desired deceleration $b$ are varied. Using steps of $\qty{0.5}{\second}$ for the reaction time up to a reaction time of $\qty{2.5}{\second}$. 
In comparison to the literature, \textcite{NHTSA2007} use a reaction time of $\qty{2}{\second}$ for the ``overwhelming majority'' of drivers. The \textcite{AASHTO2018} even uses $\qty{2.5}{\second}$ as the ``Perception-reaction time''. 

For the desired deceleration we use $\qty{-3.41}{\meter\per\second\squared}$ in \Cref{tab: collision rate 1} and $\qty{-1.71}{\meter\per\second\squared}$ in \Cref{tab: collision rate 2} for the lead vehicle.
The rear vehicle, which represents the human-driven vehicle, uses the selected naturalistic emergency deceleration of $\qty{-3.41}{\meter\per\second\squared}$.

We detect a collision between two vehicles by determining whether there is an overlap based on the positions and lengths of the vehicles.
The calculated collision rate in \Cref{tab: collision rate 1,tab: collision rate 2} is the sum of the scenarios with a collision divided by all scenarios we preprocessed, which is a total of $\sim 186000$.

\vspace{0.3cm}

\begin{table}[ht]
    \centering
    \caption{\textbf{Collision Rate - naturalistic emergency deceleration}}
    \begin{tabular}{ccc}
         \textbf{Reaction Time} & \multicolumn{2}{c}{\textbf{Collision Rate}}   \\
                                &  Sudden Braking Model & Intelligent Driver Model   \\
         \hline
         0.0&  0.03& 0.03\\
         0.5&  0.92& 0.92\\
         1.0&  8.06& 8.74\\
         1.5&  27.44& 46.65\\
         2.0& 49.04&73.09\\
         2.5& 65.13&86.09\\
         \hline
        in $\unit{\second}$ & in \unit{\percent} & in \unit{\percent} \\
    \end{tabular}
    
    \label{tab: collision rate 1}
\end{table}

\vspace{0.3cm}

\begin{table}[ht]
    \centering
    \caption{\textbf{Collision Rate - moderate deceleration}}
    \begin{tabular}{ccc}
         \textbf{Reaction Time} & \multicolumn{2}{c}{\textbf{Collision Rate}}   \\
                                &  Sudden Braking Model & Intelligent Driver Model   \\
            \hline
         0.0&  0.00& 0.00\\
         0.5&  0.01& 0.01\\
         1.0&  1.13& 1.13\\
         1.5&  6.24& 7.19\\
         2.0& 19.12&34.96\\
         2.5& 37.44&82.63\\
         \hline
        in $\unit{\second}$ & in \unit{\percent} & in \unit{\percent} \\
         
    \end{tabular}
    \label{tab: collision rate 2}
\end{table}
\subsection{Limitations}
\label{sec: limitations}
This simulation has the following limitations:
\begin{itemize}
    \item With the \emph{intelligent driver model} and our parameter selection we aim to simulate human behavior. As mentioned before, we acknowledge that this is a separate research topic and that with a more complex model there is a high probability of achieving more human-like behavior. 
    \item The \emph{intelligent driver model} itself is mainly used for traffic flow simulations and not for rear-end collisions. For this reason, we also added the \emph{sudden braking model} to show what would happen if a human driver would consistently brake with a constant maximum deceleration after a given reaction time. 
    However, this model is also limited by its simplicity and therefore cannot be considered a one-to-one simulation of human behavior.
    \item The data preprocessing pipeline to identify scenes where one vehicle is following another vehicle is based on a subjective considerations. We acknowledge that maneuver identification is also a research topic in its own right and we used a simplified approach to get the chosen initial parameters for our simulation. 
    \item The collision rate is a share of start scenes, not a real-world crash frequency. Consecutive frames of the same vehicle pair are correlated, and exposure (how often a remote vehicle is followed closely and how often the connection is lost) is not modeled. The simulation is not meant to quantify frequency, only to show that the probability is non-negligible. 
    \item Our simulation does not  take into account real-world aspects of deceleration conditions such as friction or aerodynamics or geographic variations such as road gradient and slope.
    \item Our initial parameters rely on the correctness of the \emph{uniD}~\cite{Krajewski2018, Bock2019, Krajewski2020, Moers2022} dataset and the correctness of the data recorded there. 
    Also, we only look at scenarios in the setting recorded in the \emph{uniD} dataset, which means that the vehicles following each other are limited to the road section recorded.
    \item As mentioned previously, we do not simulate lateral guidance and focus on longitudinal braking. There may be fallback strategies using lateral guidance that could perform better. We do not address those strategies in this work. 
\end{itemize}

\subsection{Results}
\label{sec: results}

This Subsection considers the simulation results from a technical perspective.
A discussion about the general context of the results can be found in \Cref{sec: discussion}.

Our results, as seen in \Cref{tab: collision rate 1} and \Cref{tab: collision rate 2}, show that there is a high collision rate given our assumptions in the simulation, and that the collision rate strongly correlates with the reaction time.

Some of the selected findings are described below:
\begin{itemize}
    \item  The overall collision rate, given the scenarios selected through preprocessing, is high, ranging up to a collision rate of $\qty{86.09}{\percent}$ when a reaction time of $\qty{2.5}{\second}$ is used.
    \item  At moderate or half deceleration, the collision rates are lower, but still result in collision rates of $\qty{19.12}{\percent}$ and $\qty{37.44}{\percent}$ for reaction times of $\qty{2}{\second}$ and $\qty{2.5}{\second}$, respectively. 
    \item  For $\qty{0.0}{\second}$ reaction time and full emergency braking for both vehicles, there is still one vehicle combination with \num{63} starting scenes in the dataset that would lead to a collision. Our hypothesis for this case is that the following vehicle did not expect the front vehicle to brake in this scenario, allowing it to go faster than the front vehicle for a short time and keeping a too short distance to the front vehicle.
    \item For comparison with the discussion of the deceleration value, we also performed the simulation with an emergency deceleration of $\qty{6.5}{\meter\per\second\squared}$ for both vehicles with the \emph{intelligent driver model} and a reaction time of $\qty{1}{\second}$. With these parameters we would have a collision rate of $\qty{10.25}{\percent}$ in our considered scenarios.
\end{itemize}

We can also see that the lower the deceleration of the front vehicle, the lower the collision rate, but our simulation only considers rear-end collisions. 
It will considerably increase the potential danger to the front if the vehicle comes to a standstill very slowly.
This means that if the deceleration is set too low, the remote-driven vehicle may decelerate with a low collision rate according to the analysis here, but because of the long stopping distance, it may slowly drive into a vehicle in front.

Returning to the considered scenario of a human-driven vehicle following a remote-driven vehicle, the results show that, given our assumptions, the fallback strategy of immediate braking to a standstill in the event of a connection loss results in a high collision rate in our simulation. 
This means that our results show that even in urban areas, if the only option for a remote-driven vehicle in the event of a connection loss is to apply a naturalistic emergency brake, we will potentially see an increase of occurrence in rear-end collisions against current urban traffic given our assumptions.

Since rear-end collisions are plausible in everyday following situations, the next logical step is to see if this hazard can be considered acceptable, which will be discussed in the next Section.


\section{Risk assessment with respect to simple fallback strategies}
\label{sec: risk assessment}

In the last Section we looked at the \emph{probability of occurrence} of such a rear-end collision when a sudden braking occurs due to a connection loss. 
As stated, our simulation can be seen here as a plausibility check to obtain a non-negligible probability of a rear-end collision occurring with the fallback strategy of braking. 

In this part we will look at the \emph{severity} of such a collision.
The overall goal is to arrive at a conceptual risk assessment for the described fallback strategy.
It makes sense to look at this case with respect to ISO~21448~\cite{ISO21448}, as it deals with hazardous vehicle behavior due to functional insufficiencies and specific triggering conditions of road vehicle systems.
Specifying a fallback strategy unable to safely deal with foreseeable operational situations can be seen as an insufficiency of specification \cite[Clause 3.12]{ISO21448}.
Therefore, we will look at the guidance process defined in the standard to examine whether the scenario considered is of relevance and potentially requires functional modifications of the remote driving system to address SOTIF-related risk.

\begin{figure*}[ht!]
    \centering
    \includegraphics[width=\linewidth,page=3]{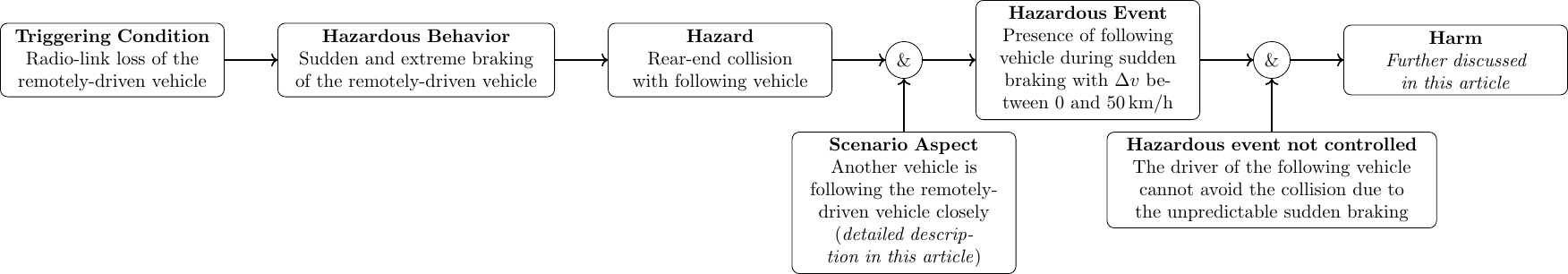}
    \caption{
    Example of hazardous event model according to ISO~21448~\cite{ISO21448}. 
    The figure shows the event chain for connection loss of a remote-driven vehicle with braking as a failure-response strategy.
    In ISO~21448~\cite[28]{ISO21448} the description of unintended braking due to a radar reflection was used as a reference. The part considering harm will be discussed in the article.
    }
    \label{fig: hazards-case}
\end{figure*}

To further illustrate the potential risk involved in the rear-end crash scenario, we will also consider what would happen if a large truck is included in the potential collision.

\subsubsection*{Why we chose this example?}
We have chosen a truck as an example here because, in the authors' view, this figurative example makes it even clearer that the potential harm is even higher when a truck is involved in this type of collision.
This again raises the question of whether the described fallback behavior can be considered acceptable when functionally specifying a remote driving system.

\subsection{Hazardous Event Model}

In \Cref{fig: hazards-case} we have applied the hazardous event model based on ISO~21448~\cite{ISO21448} to the case described. 
We will only show the relevant aspects of a scenario here and go into more detail for the whole scenario later.

\begin{itemize}
%
\item As described in the Sections before, the \emph{Triggering Condition} in this case would be the loss of the radio-link from the operator station to the remote-driven vehicle. 
%
\item This results in a \emph{Hazardous Behavior} namely a sudden and extreme braking of the remote-driven vehicle.
%
\item One \emph{Hazard} in this case would be a rear-end collision with a following vehicle. 
%
\item The important \emph{Scenario Aspect} for the selected hazard is that another human-driven vehicle is following the remote-driven vehicle closely, as only then a rear-end collision can occur. 
%
\item The \emph{Hazardous Event} is the presence of a following vehicle combined with a sudden and extreme braking of the remote-driven vehicle. For this case, as the scenario is within an urban area, the impact velocity for the rear-end collision can be up to  $\qty{50}{\kilo \meter \per \hour}$. The deceleration of the remote-driven vehicle depends on the specified fallback strategy, which  is set to an immediate sudden deceleration.
%
\item Since the driver of the following vehicle cannot \mbox{anticipate} the unpredictable sudden braking of the remote-driven vehicle, the collision \emph{is not controllable}. The driver's reaction time is too short and the stopping distance too long.
%
\item Depending on parameters, e.g. type of colliding vehicle or speed difference at impact, a \emph{Harm} to the life and health of the occupants of both vehicles can occur. This is discussed later in this Section.

\end{itemize}

\subsection{Description of the scenario}
As is common in hazard analysis, a description of the scenario is given here:

A human-driven vehicle is following a remote-driven \mbox{vehicle} on a straight road in an urban area.
The speed of the vehicles is in the range of the normal flow speed in urban road traffic in Germany.
At a given time, the remote-driven vehicle experiences a connection loss, to which it responds with a set fallback strategy.
According to the remote driving system specification, a braking maneuver is started immediately and is performed with a relatively strong deceleration to prevent the vehicle from leaving the lane or colliding with the vehicle in front.
The scenario includes an unavoidable collision between the leading, initially remote-driven vehicle and the human-driven vehicle due to a sudden braking maneuver after connection loss.
In this scenario, the weather and overall visibility conditions are good.

\subsection{Evaluation of Hazards}

The fact that sudden braking to a standstill causes unavoidable collision and thus represents \emph{Hazardous Behavior} shows that the failure-response strategy analyzed in the case study can also be interpreted as a functional insufficiency of the teleoperation system in relation to the \emph{Triggering Condition} "connection failure" within the framework of ISO~21448~\cite{ISO21448}.

With regard to ISO~21448~\cite[Clause 6.4]{ISO21448}, when looking at the \emph{severity} and \emph{controllability} of the \emph{Hazardous Event}, it should be clear that this event cannot be classified as neither ``controllable in general'' (C=0) nor ``no resulting harm'' (S=0), making it SOTIF-relevant. 

That the collision can be given at least an S1 severity rating can be seen by looking at J2980~\cite[18]{J2980}, where an S1 severity has to be assigned for frontal and rear impacts at speeds larger than $4 - \qty {10}{\kilo \meter \per \hour}$.
This is given in this scenario.

\subsection{Acceptance Criteria}

With respect to ISO~21448, the next aspect to consider is whether the specification of a sudden braking maneuver in the event of a connection loss can be considered an acceptable behavior or whether the behavior must be classified as unacceptable.

For example, ISO~21448 gives a reference to the positive risk balance, which is defined in ISO/TR~4804 as the ``benefit of sufficiently mitigating residual risk of traffic participation due to automated vehicles''~\cite{ISOTR4804:2020}.

The sudden failure of a connection, resulting in the inability to continue operation, is a new type of event whose frequency depends on the availability of the communication link in the specific operating environment.
Unless that availability can be guaranteed, it cannot be assumed to be rare.
This is in line with \textcite{Koopman2019_credible_safety}, who point out that a potential fault must not be dismissed as unrealistic without supporting data, but has to be considered in the risk analysis.

If such an event is systematically associated with vehicle behavior that leads to unavoidable accidents in normal traffic situations within the operating range of the system, possibly with serious injuries, social acceptance from a risk perspective, or approval, becomes very doubtful.

The risk balance in this kind of situation can even be seen as negative, as the introduction of remote-driven vehicles with this kind of fallback strategy would even increase the number of rear-end collisions.
Because a human driver would rarely brake this hard without a perceptible reason, these collisions are not shifted from elsewhere but are additional events introduced into road traffic.

When using an alternative risk acceptance criterion such as ALARP~\cite[Part~5, Annex~C]{IEC61508:2010}, a safety case would also have to argue why other fallback strategies, such as the ones discussed in \Cref{sec: discussion strategies}, are not reasonably practicable.

\subsection{Concrete example with increased potential harm}

The described event is generally dangerous, but if we now consider the possibility of a truck following behind, the \mbox{severity} of the collision increases significantly.

For this argumentation, we use the definition of large trucks from \cite{NHTSA2023_facts}, where they are defined as ``Trucks over 10,000 pounds GVWR [gross vehicle weight rating], including single-unit trucks and truck tractors''~\cite{NHTSA2023_facts}.

\begin{figure}[ht!]
    \centering
    \includegraphics[width=0.7\linewidth]{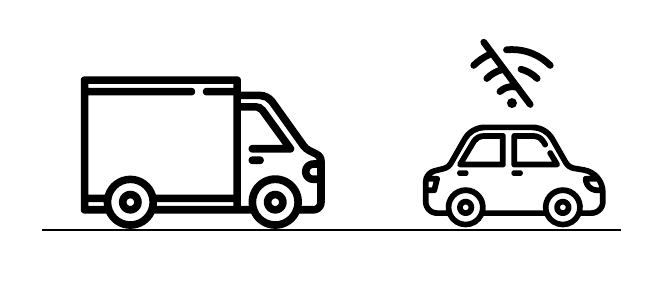}
    \caption{
  A large truck is following a remote-driven vehicle, which initiates emergency braking due to connection loss.
    }
    \label{fig: case-1}
\end{figure}

When considering a potential collision between a large truck at the rear and a remote-driven passenger vehicle at the front, a distinction can be made between the case where passengers are in the remote-driven vehicle and the case where no passengers are in the remote-driven vehicle:

\subsubsection*{Passengers inside the front vehicle}
The potential for increased harm is addressed in several accident research publications:

E.g. \citeauthor{Evans1993} state that there is a ``large influence of mass on relative driver fatality risk when two vehicles of different mass crash into each other''\cite[223]{Evans1993}.
This shows that even low-speed collisions between a truck and a passenger vehicle have a higher risk of fatality than collisions between two passenger vehicles.

That trucks involved in crashes lead to a high severity is stated in several sources \cite{Stigson2006, Chen2020, Liu2022}.

Research and publications in this area of such accidents is a vast field: reaching from looking at neck injuries for the passengers (eg. in \cite{Erbulut2014}) to collision calculations (eg. in \cite{Yang2005, Kostek2017}) to accident data (eg. in \cite{NHTSA2023_facts}).

We can conclude from this literature analysis that truck involvement increases the likelihood of significant harm to life and health in rear-end collisions.

\subsubsection*{No Passengers inside the front vehicle}
The potential for harm in the lead vehicle does not exist if there are no passengers in that vehicle.
Nevertheless, this does not change the other sources of harm of the potential collision, as the same collision could still occur here.

Looking again at ISO~21448~\cite{ISO21448}, there could still be different hazards resulting from this potential collision.
A few examples are described below:
\begin{itemize}
    \item Colliding into the rear of the remote-driven vehicle harms the truck driver.
    \item The collision pushes the vehicle onto the sidewalk, \mbox{endangering} pedestrians.
    \item The truck driver is able to react in time, but the vehicle behind the truck driver is not, resulting in a collision caused by, but not involving, the remote-driven vehicle.
\end{itemize}


\section{Discussion}
\label{sec: discussion}
In summary, we have two main findings from our analysis:
\begin{itemize}
    \item The simulation results show that rear-end collisions are plausible in everyday following situations when the remote-driven vehicle brakes as a fallback strategy.
    \item The identified SOTIF implications according to ISO~21448~\cite{ISO21448} question whether potential vehicle behavior with respect to the analyzed remote-driving fallbacks can be considered acceptable in a system design context.
\end{itemize}
We describe the potential for significant harm, even in low-speed urban environments, by including large trucks in the accident scenarios considered.

As mentioned, risk is defined as the ``combination of the probability of occurrence of harm [...] and the severity [...] of that harm [...]''~\cite{ISO26262}. 
We are able to show that the described scenario has both a non-negligible probability under plausible assumptions and a relevant potential for harm.

Notably, this scenario does not rely on a rare combination of conditions: a closely following vehicle is an everyday occurrence, and the hazardous braking is triggered by a single event — the connection loss itself. 
Its frequency therefore depends on the reliability of the connection in the intended operating environment rather than on an unusual constellation of traffic participants.

When designing a remote-driven vehicle for an urban area, taking into account a fallback strategy in the event of a connection loss could lead to design decisions other than the probably simplest one of just braking.
Unless it is revisited and replaced with an appropriate fallback strategy, braking in the event of a connection loss may be considered an inadequate specification.
Omitting this detail can lead to much higher costs once it surfaces during testing or operating the designed system.
Relying on the hope that there will never be a connection loss, or that accident scenarios will not occur or not be harmful, is not a sound basis for a safety-relevant design decision. 

For the design phase of a remote driving function in an urban area, we acknowledge that this finding probably makes the development of such a function a much more complex task:
Designing an appropriate fallback strategy is, by definition, not possible with a Level 2 function.
Specifically, for continued lateral and longitudinal guidance, the system would need to be capable of performing the dynamic driving task without human supervision.
But in the end, such system designs have to be discussed and regulated for the overall safety of all road users.
As stated in the introduction, the purpose of this article is not to solve this problem, but to point it out, in the hope of stimulating discussion about a possible specification for a solution.


\section{Conclusion and Future Research}
\label{sec: conclusion}

Despite the current emergence of remote driving in public road traffic, the question of how to deal with a connection loss, which is a major concern in the field of teleoperation, has not yet been fully addressed and can be seen as a \emph{safety blind spot}.
Depending on the intended use of the remote driving function, it is not trivial to design a fallback strategy that leads to an adequate minimal risk condition. 
This article shows that there are non-trivial problems with rear-end collisions for remote-driven vehicles when the connection is lost.

Our work also shows two things regarding the \emph{probability} and the \emph{severity} that could result from an inadequate fallback strategy:
There is (a) a non-negligible probability of collisions if the only option available to the remote-driven vehicle is sudden braking maneuver in case of a connection loss.
We have illustrated this using simple, but reasoned simulations based on a naturalistic dataset.

Also (b), in terms of \emph{severity}, we discussed the hazardous behavior showing that it is potentially unsafe and rather questionable whether such a fallback strategy can be considered acceptable. 

Furthermore, we substantiated that the potential harm of a collision caused by unpredictable deceleration is significantly higher when a large truck is involved.
In conclusion, we find that a fallback that applies naturalistic emergency deceleration on connection loss is not a proper fallback strategy in the scenarios analyzed.
We hope that this article will help developers, safety engineers, system designers, regulators, and other involved stakeholders to consider this issue in the design phase, and that it will stimulate discussion on how to deal with fallback strategies for remote-driven vehicles in urban areas. 

In future work, we will look at the problem for \mbox{vehicles} with an Automated Driving System on board, where teleoperation is used to ``extend'' the operational domain of this system. 
Another potential research question is to examine how the chosen models behave on datasets other than the \emph{uniD} dataset. 
Possible interactions in front of the remote-driven vehicle could also be considered depending on real datasets. 
In addition to considering the rear of the remote-driven vehicle, a follow-up question would be to assess the collision potential of vehicles in front of the remotely operated vehicle.

\fi   


\ifabstractonly\else   
\ifanonymous
\vspace{2em}
\textit{Funding and CRediT are anonymized.}
\vspace{2em}
\else

\section*{Funding}
This work was partly supported by the German Federal Ministry for Economic Affairs and Energy within the project ``Automatisierter Transport zwischen Logistikzentren auf Schnellstraßen im Level~4 (ATLAS-L4)'' under grant number 19A21048H.
The funding source had no involvement in the study design, in the analysis and interpretation of the concepts presented, in the writing of the article, or in the decision to submit it for publication.

\section*{CRediT authorship contribution statement}
\textbf{Leon Johann Brettin:} Conceptualization, Methodology, Investigation, Visualization, Software, Formal analysis, Data curation, Writing -- original draft, Writing -- review \& editing.
\textbf{Niklas Braun:} Conceptualization, Methodology, Investigation, Software, Formal analysis, Data curation, Writing -- review \& editing.
\textbf{Robert Graubohm:} Writing -- review \& editing.
\textbf{Markus Maurer:} Supervision, Writing -- review \& editing.

\fi

\section*{Data availability}
The simulation is based on the \emph{uniD} dataset~\cite{Krajewski2018, Bock2019, Krajewski2020, Moers2022}, which is available for non-commercial research upon request from its provider (\url{https://levelxdata.com/unid-dataset/}). The authors are not permitted to redistribute it.

\section*{Declaration of generative AI and AI-assisted technologies in the manuscript preparation process}
As the authors of this article are not native English speakers, the writing of this article was assisted by AI translation and spelling assistance tools, namely DeepL and Claude.
\fi   

\ifmaintext   
\renewcommand*{\bibfont}{\footnotesize}

\printbibliography
\fi

\end{document}